\documentclass[fleqn,usenatbib]{mnras}

\usepackage[utf8]{inputenc}

\usepackage{acronym}
\usepackage{ae,aecompl}
\usepackage{booktabs}
\usepackage{dcolumn}
\usepackage{graphicx}
\usepackage{amsmath,amsfonts,amssymb}
\usepackage{mathrsfs}
\usepackage[normalem]{ulem}

\usepackage[T1]{fontenc}

\let\oldhref\href
\renewcommand{\href}[2]{\oldhref{#1}{\hbox{#2}}}

\title{Three-Dimensional Supernova Explosion Simulations of\phantom{ } 9-, 10-, 11-, 12-, and 13-M$_{\odot}$ Stars}

\author[A. Burrows et al.]{Adam Burrows$^{1}$\thanks{E-mail: aburrows@princeton.edu},
David Radice$^{1,2}$ ,
\newauthor
David Vartanyan$^{1}$\\
\\
$^{1}$Department of Astrophysical Sciences, Princeton University, Princeton, NJ 08544\\
$^{2}$ Institute for Advanced Study, 1 Einstein Dr, Princeton NJ 08540\\
}

\begin{document}
\maketitle

\begin{abstract}

Using the new state-of-the-art core-collapse supernova (CCSN) code F{\sc{ornax}}, we
have simulated the three-dimensional dynamical evolution of the cores of
9-, 10-, 11-, 12-, and 13-M$_{\odot}$ stars from the onset of collapse.
Stars from 8-M$_{\odot}$ to 13-M$_{\odot}$ constitute roughly 50\% of all massive stars,
so the explosive potential for this mass range is important to
the overall theory of CCSNe.  We find that the 9-, 10-, 11-, and 12-M$_{\odot}$ models
explode in 3D easily, but that the 13-M$_{\odot}$ model does not.  From these findings,
and the fact that slightly more massive progenitors seem to explode \citep{vartanyan2019}, we suggest
that there is a gap in explodability near 12-M$_{\odot}$ to 14-M$_{\odot}$ for
non-rotating progenitor stars.  Factors conducive to explosion
are turbulence behind the stalled shock, energy transfer due to neutrino-matter absorption and
neutrino-matter scattering, many-body corrections to the neutrino-nucleon scattering rate,
and the presence of a sharp silicon-oxygen interface in the progenitor.
Our 3D exploding models frequently have a dipolar structure, with the two asymmetrical
exploding lobes separated by a pinched waist where matter temporarily continues
to accrete.  This process maintains the driving neutrino luminosty, while partially
shunting matter out of the way of the expanding lobes, thereby modestly facilitating
explosion. The morphology of all 3D explosions is characterized by multiple bubble
structures with a range of low-order harmonic modes.  Though much remains to be
done in CCSN theory, these and other results in the literature suggest that, at
least for these lower-mass progenitors, supernova theory is converging on a
credible solution.

\end{abstract}

\begin{keywords}
stars - supernovae - general
\end{keywords}

\section{Introduction}
\label{introduction}

\begin{figure*}
\center{\includegraphics[width=0.7\textwidth]{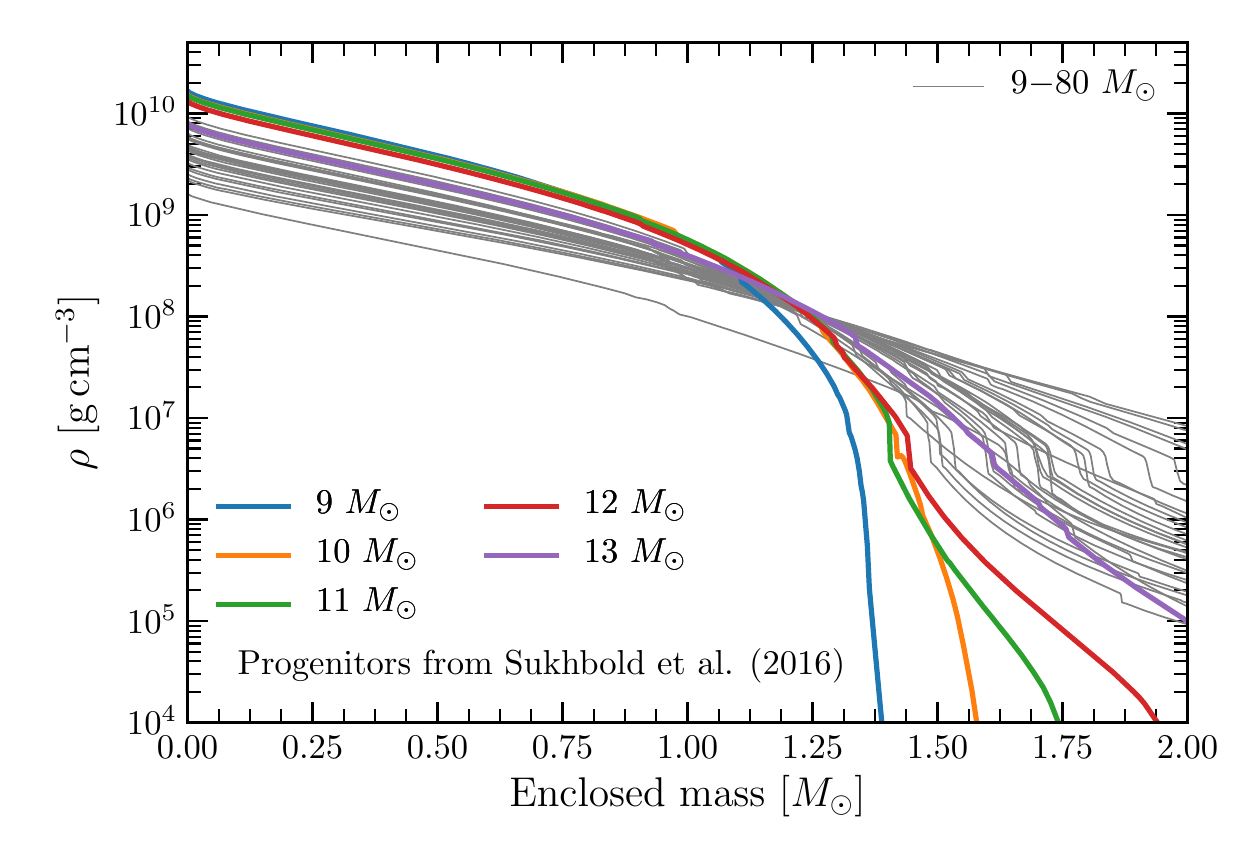}}
\caption{The mass-density ($\rho$, in g cm$^{-3}$) versus interior mass (in units of M$_{\odot}$)
for various representative progenitor models from \protect\cite{sukhbold:16}. The profiles for the 9-, 10-, 11-, 12-, 
and 13-M$_{\odot}$ models are highlighted in color. Comparisons between these and the other 
profiles (in gray) up to 80-M$_{\odot}$ put this lower-mass subclass into the larger context 
of progenitor initial models. Note that the 13-M$_{\odot}$ model is distinct from the others
highlighted in this low-mass progenitor study.  See the text for a discussion.
}
\label{profiles}
\end{figure*}

Approximately $\sim$50\% of the mass function of massive stars above $\sim$8.0 M$_{\odot}$
lies below $\sim$13.0 M$_{\odot}$. Since only stars more massive than $\sim$8.0 M$_{\odot}$ 
can end their lives as core-collapse supernovae (CCSNe) (simultaneously giving birth to either
neutron stars or black holes), understanding the mechanism and character of supernova 
explosions (if they occur) in this modest mass range, assumes an outsized astrophysical importance.
Traditionally, those who model the core-collapse and explosion phases of massive stars inherit
progenitor models at the cusp of core collapse from experts in massive star evolution.
The latter simulate a star's passage through successive burning phases until an unstable
Chandrasekhar white dwarf core emerges at the star's center, at which point the physical profiles
of that core are mapped onto the grid of a supernova code to carry the dynamical,
oftimes multi-dimensional, evolution forward.  Aside from the stochasticity and chaos
associated with the turbulence that attends both progenitor and supernova convective 
instabilities, the structure of the roughly spherical ``initial" model determines the outcome
of the supernova simulation, and, it is hoped, the outcome of stellar death.  In particular,
a progenitor's radial mass density profile seems to determine much of the subsequent explosive behavior.  
Figure \ref{profiles} portrays a representative collection of such profiles.
It has been observed, indeed with quantitative variations from modeler to modeler and with some
degree of non-monotonic behavior with progenitor ZAMS mass \citep{wh07,sukhbold:16,sukhbold2018},
that massive stars at the lower end of the mass function have steeper
mass density profiles with radius than those at the higher end. 

It is thought that such steep profiles result in cores that explode 
easily by the proto-neutron star (PNS) neutrino-driven wind mechanism 
\citep{burrows1987_wind,burrows:95}, even in one-dimension (1D, spherical), and this has 
been shown to be the case \citep{kitaura,fischer:10,radice:17} for the
pioneering 8.8-M$_{\odot}$ model of Nomoto \citep{nomoto:84,nomoto:87}. However, such ``electron-capture" 
supernovae (ECSNe) occupy a problematic region of model space \citep{woosley_heger:2015}, 
one in which burning under electron-degenerate conditions with subsequent flashes could 1) eject envelope 
matter before collapse, 2) radically restructure the core, or 3) compromise the accuracy of 1D
stellar evolution simulations. Nevertheless, spherical progenitor models with masses from 
8.1 M$_{\odot}$ to 9.6 M$_{\odot}$,\footnote{The 8.1 M$_{\odot}$ and 9.6 M$_{\odot}$ models
were for 10$^{-4}$ and zero metallicity, respectively.} and steep density profiles in 
the outer Chandrasekhar mantle, have exploded in 1D CCSN simulations 
\citep{kitaura,fischer:10,muller:2012,melson:15a,radice:17}. These supernova models universally involve 
low explosion energies ($\sim$10$^{50}$ ergs $\equiv$ 0.1 Bethe). When performed
in 2D \citep{burrows2007_onemg,muller:2012,radice:17} or 3D (for the 9.6 M$_{\odot}$ model; \cite{melson:15a}), 
the explosions are not only low-energy, but quasi-spherical, and such explosions are 
likely to yield low-mass neutron stars with low-velocity neutron star kicks. A reasonable
conclusion is that if the progenitor mass density profiles are as steep as found 
in these models (Figure \ref{profiles}), the theory and rough explosion numbers arrived at using modern 
supernova codes that incorporate neutrino transport and heating may be roughly reproducing 
Nature.

However, the 1D and 2D supernova models of \cite{radice:17} for the 9, 10, 11 M$_{\odot}$ progenitor models 
of \cite{sukhbold:16}, though spanning a low-mass segment of progenitor parameter space and manifesting
a monotonic sequence in mass-density profile from very steep to progressively less steep, 
do not behave monotonically, nor do they all explode easily.  The 9-M$_{\odot}$ model explodes easily in 
2D by a neutrino-driven wind mechanism (though not in 1D), but the 10-M$_{\odot}$ model does not explode 
in 1D or 2D without significant progenitor velocity perturbations
\citep{CoOt13,muller_janka_pert,mueller:17,abdik:16,takahashi:16}.  On the other hand, 
as with the 9-M$_{\odot}$ model, the 11-M$_{\odot}$ model does not explode in 1D, but does explode in 2D, 
and without the aid of perturbations. Moreover, the 12-M$_{\odot}$ model of \cite{wh07}, simulated in 2D by 
\cite{burrows:18}, \cite{vartanyan2018a} and \cite{OcCo18}, does not explode at all unless aided 
by such things as rotation or significant velocity perturbations \citep{vartanyan2018a}, but 
the 16-M$_{\odot}$ progenitor model from \cite{wh07} with a significantly shallower mass density 
profile explodes easily in 2D and 3D \citep{vartanyan2019}.  Collectively, this behavior with 
variations in mass density profile along the progenitor mass continuum, for which in this mass 
range the ``compactness" parameter \citep{oconnor_ott:11,oconnor_ott:2013,ott2018_rel} is monotonic 
demonstrates once again that the compactness parameter is not predictive of ``explodability" \citep{burrows:18}. 

Supernova modeling experience
in the recent past now suggests that as the progenitor mass density profile shallows  
explodability by the wind mechanism is at first easy, and then difficult or impossible. 
However, with still shallower profiles, explosions by a convection-aided, neutrino-driven 
mechanism become easier again, though this is not a classic wind. For these 
progenitors, the post-bounce, pre-explosion mass accretion rates are large 
enough to result in larger mass densities between the stalled shock wave and 
PNS core. The correspondingly larger neutrino absorption optical depths in this mantle region  
lead to more efficient neutrino heating that, with the aid of multi-dimensional 
effects \citep{herant1994,burrows:95,murphy:08}, ignites an explosion. These massive stars have the potential 
to yield larger explosion energies, with larger kick speeds and neutron star masses. 
Such explosions are also aided by the accretion of steep silicon/oxygen interfaces 
\citep{vartanyan2018a}.  The associated discontinuity in the mass accretion rate
through the shock results in a decrease in the inhibiting tamp, but there is a delay
in the corresponding decrease in the neutrino luminosity due to the finite advection 
time between the shock and the inner core.  By accompanying a temporarily unaltered neutrino 
luminosity with an immediate decrease in the ram pressure, the shock wave can be 
kicked into explosion by suddenly achieving the critical condition \citep{goshy}.  
For progenitors with initial mass density profiles intermediate between those of these 
exploding classes, post-bounce accretion smothers the wind, while failing 
to provide a large enough neutrino optical depth in the mantle.  To date, for 
these progenitor structures, currently residing near $\sim$12 M$_{\odot}$ 
\citep{wh07,sukhbold:16,sukhbold2018}, 1D and 2D supernova models do not 
explode in the context of the default physics employed, carried out for as 
long as one second post-bounce, and without initial rotation nor substantial 
progenitor seed perturbations. However, we reiterate that the progenitor profiles 
for massive stars have not theoretically converged and, moreover, 
that the mapping between ZAMS mass and progenitor structure has not 
been definitively settled \citep{wh07,sukhbold:16,sukhbold2018,meakin:11,CoChAr15,muller_viallet:16,mueller:17}.

In this paper, using our new supernova code F{\sc{ornax}} (see \S\ref{methods} and 
\cite{fornaxcode:18}, we simulate in three spatial dimensions (3D) the self-consistent 
behavior at the lower-mass end (9-, 10-, 11-, 12-, and 13-M$_{\odot}$) of a suite of 
supernova progenitors from \cite{sukhbold:16}. This work is in part a 3D extension 
of our earlier 1D and 2D study \citep{radice:17}.\footnote{F{\sc{ornax}} has 
also been employed for a variety of 1D and 2D supernova simulations 
\citep{wallace:16,skinner2016,radice:17,vartanyan2018a,burrows:18}, 
and recently to follow a 3D CCSN explosion for a 16-M$_{\odot}$ progenitor 
$\sim$one-second post-bounce \citep{vartanyan2019}.}  As stated earlier, 
these masses span a large fraction of the mass function. 

An earlier generation of pioneering 3D simulations 
\citep{hanke:13,Tamborra14,couch:14,melson:15a,melson:15b,lentz:15,bmuller_2015,takiwaki2016,roberts:16,ott2018_rel,oconnor_couch2018,summa:18,glas2018}, employing various necessary
simplifications, have set the stage for this study, but 3D simulations are still rare. 
Moreover, some earlier work did not include
inelastic scattering nor velocity-dependent transport \citep{couch:14,ott2018_rel,roberts:16},
did not include inelastic scattering on nucleons \citep{glas2018}, employed ``ray-by-ray+" dimensional 
reduction \citep{hanke:13,Tamborra14,lentz:15,bmuller_2015,melson:15a,melson:15b,takiwaki2016,summa:18},
cut out the inner core, or performed the simulations in the inner core in 1D.  Depending upon
how large this inner 1D core was, the latter procedure could suppress, partially 
or in full, PNS convection \citep{dessart_06,radice:17,glas2018b}. Such convection
has been shown to facilitate the explosion of some models \citep{radice:17}.

Though still expensive, our 3D F{\sc{ornax}} runs avoid all these issues, 
and are state-of-the-art, but still compromise on aspects of the problem.  In particular, 
we employ the two-moment closure M1 method \citep{vaytet:11,fornaxcode:18}, which is not 
multi-angle.\footnote{No such capability with full physics that can follow the evolution for a 
physically relevant timescale currently exists.} In addition, our energy-group number and spatial resolution
should be increased,\footnote{We use 12 groups per neutrino species and our $r\times\theta\times\phi$ 
spatial discretization is $678\times128\times256$.} perhaps significantly, to 
ensure the results are converged.\footnote{However, the resolution dependence of
all extant 3D runs remains a universal issue for the CCSN modeling community and
our energy and spatial resolutions are better than or comparable to those of most studies in the 3D literature.}
Furthermore, we use approximate general relativity (GR) to address the enhanced attractive 
strength of GR and gravitational redshifts, and not full GR \citep{roberts:16,ott2018_rel}.
Finally, we lump the $\nu_{\mu}$, $\bar{\nu}_{\mu}$, $\nu_{\tau}$, and $\bar{\nu}_{\tau}$ neutrinos into one
species.  It has been suggested that this approach can have its limitations \citep{2017PhRvL.119x2702B}.
Nevertheless, F{\sc{ornax}} belongs to a new generation of CCSN simulation 
codes now emerging with 3D radiation/hydrodynamic capabilities that, though
resource-intensive and perforce incorporating various 
and sundry approximations, finally promise to address the core-collapse 
supernova phenomenon with full physical fidelity.

In \S\ref{methods}, we briefly summarize our computational setup and the 
microphysics incorporated into F{\sc{ornax}}.  Then, in \S\ref{basics}
we present the central results and derived quantities of our 3D simulations.
Following this, in \S\ref{debris} we discuss the characteristics of the explosion debris field
and morphology, and then in \S\ref{conclusions} we recap our most 
important findings, summarize the current status of supernova theory, 
and speculate on productive future directions.

\section{Methods and Setup}
\label{methods}

All the 3D and 2D runs performed in this paper used the multi-group radiation/hydrodynamics 
code F{\sc{ornax}}.  To date, results using this new capability have been published in
\cite{skinner2016}, \cite{wallace:16}, \cite{radice:17}, \cite{vartanyan2018a},
\cite{burrows:18}, \cite{fornaxcode:18}, and \cite{vartanyan2019}.
The methodologies and equations solved are described in detail 
in \cite{fornaxcode:18}. The full suite of microphysics and the approach to 
approximate general relativity (GR) employed are summarized in \cite{vartanyan2019}, in particular
in Appendices A and B, respectively, in that paper. We restrict the calculation of 
gravity to the monopole term, corrected for approximate GR \citep{marek:06} and 
including gravitational redshifts of the neutrino spectra. The comoving-frame transport is a multi-dimensional
variant of the two-moment scheme M1 \citep{vaytet:11}, with analytic closures for the second 
and third radiation moments \citep{fornaxcode:18}. Weak magnetism and recoil effects
are handled using the formulae in \cite{horowitz2002} without truncation. Velocity-dependent transport terms
to order $(v/c)$ are included, inelastic energy redistribution for neutrino-electron 
and neutrino-nucleon scattering are handled using the method of \cite{2003ApJ...592..434T}
and \cite{burrows_thompson2004}, and detailed neutrino-matter scattering and absorption opacities are
calculated using the formulations of \cite{burrows:06}, augmented to include
the many-body correction to the neutrino-nucleon scattering rates of \cite{horowitz:17}.
The latter progressively decreases these neutral-current scattering opacities with increasing density
in a manner that approximately captures the associated many-body physics \citep{1998PhRvC..58..554B}.
The effects of this term on core-collapse physics and phenomenology are explored 
in \cite{burrows:18} and \cite{vartanyan2018a}.  

We simulated the multi-D transport of the electron-type neutrinos ($\nu_e$),
anti-electron-type neutrinos ($\bar{\nu}_e$), and ``$\nu_{\mu}$"s, where the
latter bundles the $\nu_{\mu}$, $\bar{\nu}_{\mu}$, $\nu_{\tau}$, and $\bar{\nu}_{\tau}$
neutrinos. In 3D, we used twelve energy groups for each species, spanning 1 to 300 MeV
for the $\nu_e$s and 1 to 100 MeV for the others and in 2D we used twenty energy
groups over the same ranges. 

The calculations were all performed using the same spherical, dendritic grid \citep{fornaxcode:18}, 
which deresolves the angular zoning as the center is approached to roughly maintain 
the same aspect ratio of the spatial zones in the inner $\sim$20 kilometers and to minimize
the Courant penalty of spherical convergence.  The zoning was 678$\times$128$\times$256 
($r\times\theta\times\phi$) in 3D and 678$\times$256 ($r\times\theta$) in 2D.  
The inner zone had a $\Delta{r}$ of 0.5 kilometers, and the radial zoning was 
roughly constant in the inner $\sim$20 km and logarithmic exterior to $\sim$20 
kilometers out to an outer spatial boundary of 20,000 kilometers.  
The 3D runs were begun 10 milliseconds after bounce and were mapped from 
a corresponding 1D run and continued forward in 3D.

Though 9, 10, and 11 M$_{\odot}$ simulations in \cite{radice:17} employed the
non-rotating progenitors from \cite{sukhbold:16}, as we do here for all our models (see Figure \ref{profiles}), the EOS
used in that study was that of \cite{lattimer:91} (LS220).  Because the SFHo \citep{steiner:13} EOS
is still consistent with all known nuclear and astrophysical constraints
\citep{2017ApJ...848..105T}, whereas the LS220 EOS no longer is, we have opted to use
the SFHo EOS in this study. Therefore, we have redone for this paper the comparison
2D simulations for the 9, 10, and 11 M$_{\odot}$ progenitors using the SFHo EOS.
Also, whereas \cite{radice:17} investigated the effects of the many-body
corrections of \cite{horowitz:17} (see also \cite{burrows:18} and \cite{1998PhRvC..58..554B})
to the neutral-current scattering rates off free nucleons, we here include
such corrections by default. 

A number of authors have touted
the potential role of perturbations due to pre-supernova turbulence
\citep{CoOt13,muller_janka_pert,mueller:17,abdik:16,takahashi:16},
preliminary 3D progenitor models are now emerging \citep{CoChAr15,mueller:17},
and \cite{radice:17} and \cite{vartanyan2018a} have compared various models with and without perturbations.
For our 3D runs, we imposed slight $\ell = 10$/$m=1$/$n = 4$ velocity perturbations on the initial 
3D model between 200 and 1000 kilometers, using the scheme of \cite{muller_janka_pert}, 
as implemented by \cite{radice:17}. However, the magnitude of these perturbations, 
meant to seed instabilities that might arise, was only 100 km s$^{-1}$.  This 
is to be compared with the radial speed of tens of thousands of km s$^{-1}$ at 
the beginning of the 3D phase. This prescription imposes a minimal 
degree of perturbation merely to seed turbulence
and seems irrelevant to the viability of the explosions we witness.
We impose the same perturbations for the comparison 2D simulations, 
evaluating them in this case at $\phi = 0$. Hence, it is highly unlikely that these perturbations 
played any substantive role in the subsequent growth of turbulence, 
and were likely dwarfed in effect by the ``grid noise."

This paper is meant in part to be a three-dimensional continuation of the
2D and 1D study in \cite{radice:17}.  However, that paper explored
9, 10, 11, 8.1, 8.8, and 9.6 M$_{\odot}$ models, while we here focus on
the 9, 10, 11, 12, and 13 M$_{\odot}$ models of \cite{sukhbold:16}.  Since the 8.1 and 9.6 M$_{\odot}$
progenitors investigated by \cite{radice:17} were for $10^{-4}$ and zero metallicity
and our goals here are to understand the common core-collapse supernova phenomenon,
we drop these models from further consideration.  In addition, since there
is no model in \cite{sukhbold:16} that corresponds to the 8.8 M$_{\odot}$ 
progenitor of Nomoto \citep{nomoto:84} or that is lower in mass than 9.0 M$_{\odot}$, 
the latter mass serves as the lower mass limit of this study.

\section{Basic Explosion Results}
\label{basics}

Tables \ref{sn_tab} and \ref{sn_tab2} provide some useful summary numbers
associated with our 3D and 2D core-collapse simulations.  For Table \ref{sn_tab},
the explosion energy, the baryon and gravitational masses of the residual proto-neutron star,
the average PNS radius (defined as the average radius interior to an isodensity
surface at 10$^{11}$ g cm$^{-3}$), and the envelope binding energy of the off-grid material
exterior to 20,000 kilometers (km) are given at t(final), the time post-bounce at the end 
of each simulation.  All the models, except the 13-M$_{\odot}$ model, explode.  Table \ref{sn_tab2}
lists the mean shock radii and speeds at the end of each simulation. In 3D, the former range from $\sim$15,200 
km to $\sim$2,600 km. Since all the exploding models have diagnostic energies (see below) at the
end of the calculation in excess of the off-grid overburden energy (see Table \ref{sn_tab}), these
numbers should continue to increase. 

\begin{figure*}
\center{\includegraphics[width=0.7\textwidth]{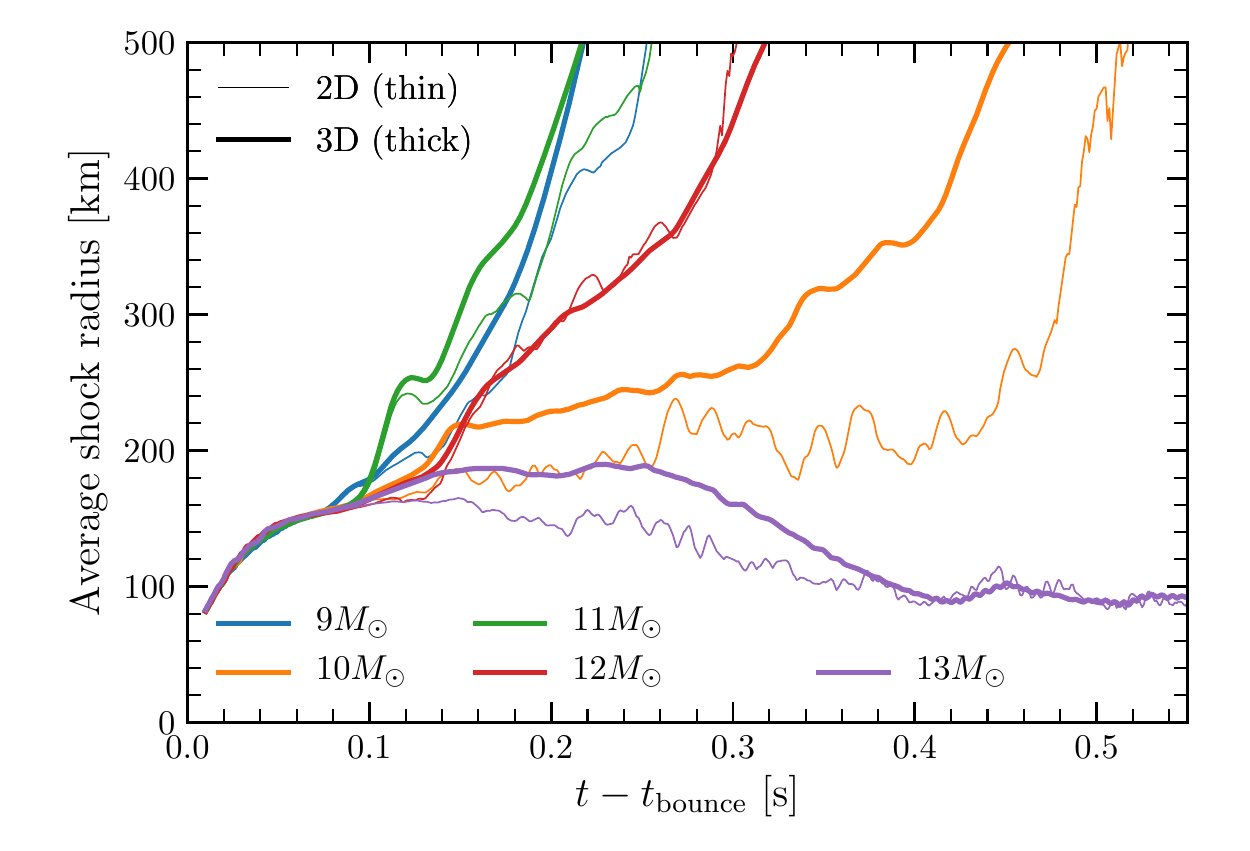}}
\caption{The solid-angle-weighted average shock wave radius (in kilometers) versus time after bounce 
(in seconds) for the 9-, 10-, 11-, 12-, and 13-M$_{\odot}$ models of this study in 3D (thick) 
and 2D (thin).  All the 3D and 2D models, except the 13-M$_{\odot}$ model, explode. Shown are
the radii until 0.55 seconds after bounce, though the runs were frequently carried out further 
(see Table \ref{sn_tab}).  The 9-M$_{\odot}$ and 11-M$_{\odot}$ models explode within $\sim$100
milliseconds of bounce, the 12-M$_{\odot}$ progenitor requires $\sim$40 milliseconds longer, while the 
10-M$_{\odot}$ model is clearly exploding by $\sim$0.3 and 0.45 seconds in 3D and 2D, respectively.
Generally, the 3D models explode slightly earlier than the 2D models, though for the 12-M$_{\odot}$
progenitor the 3D and 2D models are launched at roughly the same time.  We note that the 13-M$_{\odot}$ model
in this progenitor model suite not only does not explode in either 2D or 3D, but that it has 
a muted silicon/oxygen interface jump in density (and entropy) relative to that of the 
others (see Figure \ref{profiles}) that resides further out in interior mass.
These factors seem to have an impact on the ``explodability" of that core.
Moreover, in \protect\cite{burrows:18} and \protect\cite{vartanyan2018a}, the 2D 12-M$_{\odot}$ model, using default physics,
did not explode, but this initial model was from a different progenitor
suite \protect\citep{wh07} for which the 
12-M$_{\odot}$ model does not have as pronounced a silicon/oxygen density discontinuity. See the text for
a discussion of these trends.
}
\label{shock}
\end{figure*}

Figure \ref{shock} depicts the first 0.55 seconds after bounce of the evolution of 
the mean shock radius for both 3D (thick) and 2D (thin) models.\footnote{Given that 3D is the proper
context for simulations in Nature, we emphasize that we have carried and calculated the 
2D models in this paper merely for important technical comparison. As an aside, we comment that none of 
these models explodes in 1D.} Except for the 13-M$_{\odot}$ model, all the 3D and 2D models explode.  
As indicated on this figure, all the models evolve similarly in the first $\sim$100 milliseconds.
This is a consequence of the characteristic timescale in this study for convection and turbulence
to kick in and reflects the very small initial seed perturbations imposed (see \S\ref{methods}).
It is in part the breaking of symmetry due to the onset of turbulence, with its associated dynamics 
and turbulent stresses, that enables the divergence of the behaviors of the various stalled shocks.
The mean shock radii of the 9-M$_{\odot}$ and 11-M$_{\odot}$ models never really stall or recede 
and these models explode within $\sim$100 milliseconds of bounce.  As with many of our models, the 
9-M$_{\odot}$ core explodes immediately upon accretion through the partially stalled shock of its sharp
silicon/oxygen interface \citep{vartanyan2018a}. The core of the 12-M$_{\odot}$ progenitor 
requires $\sim$40 milliseconds longer to launch its explosion, while the
10-M$_{\odot}$ model takes rather much longer to supernova, but is clearly doing so 
by $\sim$0.3 and $\sim$0.45 seconds in 3D and 2D, respectively.
Generally, the 3D models explode slightly earlier than the 2D models, though 
for the 12-M$_{\odot}$ progenitor the 3D and 2D models are launched at roughly the 
same time.  This behavior is consonant with that found by \cite{vartanyan2019} for the 16-M$_{\odot}$
star from \cite{wh07}, who observed that for this progenitor the 3D explosion 
preceded the 2D explosion by $\sim$50 milliseconds.  We note that using their FMD 
(fully-multi-dimensional) code AENUS-ALCAR, which did not incorporate redistribution by inelastic scattering nor
the many-body correction to neutral-current neutrino-nucleon scattering \citep{horowitz:17},
\cite{glas2018} also found that their 3D model exploded before their 2D model, though it exploded
at $\sim$300 ms after bounce, later than we find. This difference could easily reflect differences in
numerical approach and microphysics. Why the 3D models explode earlier than the 2D models
is unclear, but was anticipated in \cite{dolence:13}.\footnote{See also \cite{hanke:12} 
and \cite{hanke:13} for an alternate view.} 

As stated earlier, the 13-M$_{\odot}$ model in this progenitor suite does 
not explode in either 2D or 3D. As Figure \ref{profiles} demonstrates, this model not only has
a shallower density profile (and, therefore higher post-bounce mass accretion rates), but
a more muted density jump at its silicon/oxygen interface relative to that 
of the other ``low-mass" progenitors highlighted in this paper.  This interface 
also resides further out in interior mass. These factors clearly have a bearing on
the outcome we find.  Due to its steep density profile, the 9-M$_{\odot}$ model 
transitions rather quickly to a neutrino-driven wind explosion \citep{burrows1987_wind} 
and achieves a low asymptotic explosion energy near 0.1 Bethe.  The 10-M$_{\odot}$,
11-M$_{\odot}$, and 12-M$_{\odot}$ models explode similarly weakly (though quickly), due in part to the 
low absorbing mass (hence, low neutrino optical depth) in the gain region 
\citep{1985ApJ...295...14B} and the low driving $\nu_e$ and $\bar{\nu}_e$ neutrino 
luminosities.\footnote{Note again that these models also have 
more prominent Si/O interfaces.} Both the low luminosities (which are powered 
in this early phase mostly by accretion) and low neutrino optical 
depths are consequences of the steep core mass density profiles of these 9-M$_{\odot}$,
10-M$_{\odot}$, 11-M$_{\odot}$, and 12-M$_{\odot}$ models. However, it seems that 
for these models the positive effect of post-shock turbulence and neutrino heating outweigh 
the negative effects of the accretion ram and the critical condition for explosion 
\citep{goshy,murphy:08} can be met. But as the mass density shallows and if the Si/O interface
is more subtle, we witness a decreasing tendency to explode, as manifest by our result
for the 13-M$_{\odot}$ progenitor. However, as our positive result for the corresponding
16-M$_{\odot}$ progenitor \citep{vartanyan2019} demonstrates, explodability returns despite a shallower, higher 
density mantle.  For such models the neutrino optical depth and accretion-powered 
luminosities are more significant and, particularly if the Si/O interface jump is
not small, ``explodability" seems to return.  Therefore, we are seeing a gap in explodability,
near 13-M$_{\odot}$ for the \cite{sukhbold:16} progenitor models and our F{\sc{ornax}} implementation 
(\S\ref{methods}).  We saw a similar gap at 12-M$_{\odot}$ in \cite{burrows:18} 
and \cite{vartanyan2018a} for the 2D models of those studies with a different 
progenitor suite \citep{wh07}, as did \cite{OcCo18}.  Hence, it may well be that 
1) the lowest-mass progenitors experience low-energy (one to a few$\times$10$^{50}$ ergs?)
explosions that are partially wind-like; 2) higher-mass progenitors experience 
higher-energy explosions due to higher driving luminosities and neutrino absorption 
optical depths in the gain region, despite the higher accretion ram tamp; and 3)
there is a gap in explodability.\footnote{We still suspect that for the shallowest
density profiles and highest outer envelope binding energies the latter may be 
too much to overcome \citep{burrows:18}.} It may be that some flavor of rotation, or new/better
physics and numerics, could close this gap, but currently the presence of a gap
in explodability, as indicated by our 3D calculations to date, seems plausible. 
We emphasize that the ``compactness parameter" \citep{oconnor_ott:11} of the 
9-M$_{\odot}$ through our 16-M$_{\odot}$ \citep{vartanyan2019} models we have 
studied to date is monotonic with progenitor mass, but that the outcomes are not.\footnote{For
the 9-, 10-, 11-, 12-, and 13-M$_{\odot}$ stars of \cite{sukhbold:16}, the compactness 
parameters, all small, are 3.83$\times 10^{-5}$, 2.64$\times 10^{-4}$, 7.67$\times 10^{-3}$, 
2.23$\times 10^{-2}$, and 5.93$\times 10^{-2}$, respectively.}

\begin{figure*}
\center{\includegraphics[width=0.7\textwidth]{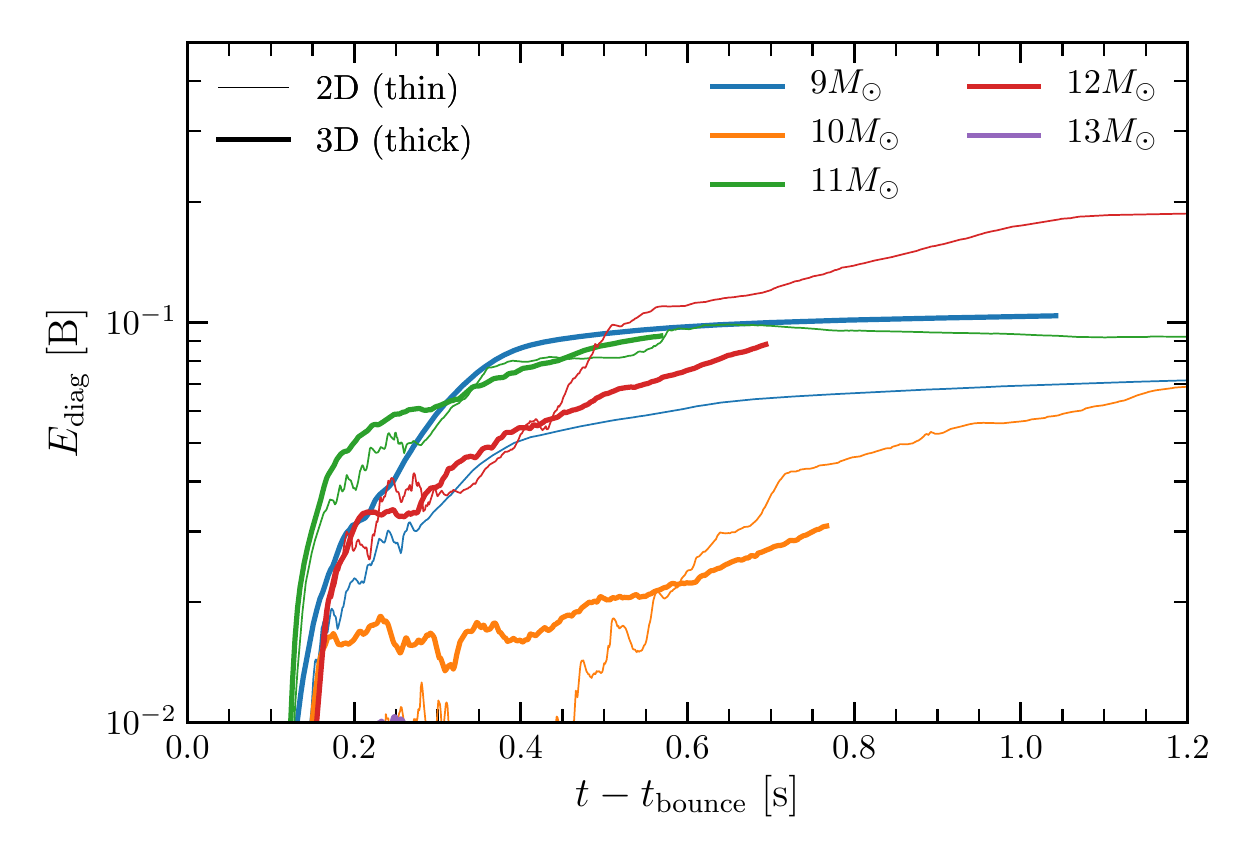}}
\caption{Diagnostic explosion energies (in Bethes, $\equiv$ 10$^{51}$ ergs)
versus time after bounce (in seconds) for the 9-, 10-, 11-, and 12-M$_{\odot}$ models
calculated for this investigation.  Note that since the 13-M$_{\odot}$ model
does not explode in either 3D or 2D it does not register as one of the 
plotted lines (it would have been purple). The thick lines are for the 3D models
and the thin lines are for the 2D models.  The diagnostic energy includes the sum
of the gravitational, kinetic, thermal, and recombination energies of the ejecta, 
but not the binding energy of the off-grid matter exterior to our 20,000-kilometer
outer boundary.  However, as can be seen in Table \ref{sn_tab}, the outer envelope binding 
energies for each of these low-mass models are quite small.  We note that the 9-M$_{\odot}$ models have 
total explosion energies (in 3D and 2D) that have almost asymptoted to their final values.  
See text for a discussion.
}
\label{energy}
\end{figure*}

Figure \ref{energy} portrays the ``diagnostic" explosion energy \citep{melson:15a} 
evolution for our 9-, 10-, 11-, and 12-M$_{\odot}$ 3D and 2D models.  Since the 
13-M$_{\odot}$ model does not explode in either 3D or 2D, it does not appear as one 
of the plotted lines.  By diagnostic energy, we mean the sum of the gravitational, 
kinetic, thermal, and recombination energies of the ejecta. The ejecta are defined
as matter that instantaneously appears unbound by the standard Bernoulli condition,
and the diagnostic energy does not include the binding energy of the matter 
exterior to our 20,000-kilometer outer boundary. The latter is given in Table 
\ref{sn_tab} and we see that in all exploding cases the diagnostic energy has 
overcome the mantle binding term by the end of all our simulations, and is still 
climbing for most.  In fact, the total asymptotic explosion energy (Table \ref{sn_tab}) 
is indeed positive for all our exploding models, though those energies are still 
modest and seem destined to be no more than $\sim$one to a few$\times$10$^{50}$ ergs.  
However, for this low-mass massive star subset, this may be realistic 
\citep{morozova_energy}.  The 9-M$_{\odot}$ models have total explosion energies 
(in 3D and 2D) that have almost asymptoted.  We note that our 3D 9-M$_{\odot}$ 
model is one of the only state-of-the-art 3D models in the literature to have approached an
asymptotic value.

\begin{figure*}
\center{\includegraphics[width=0.7\textwidth]{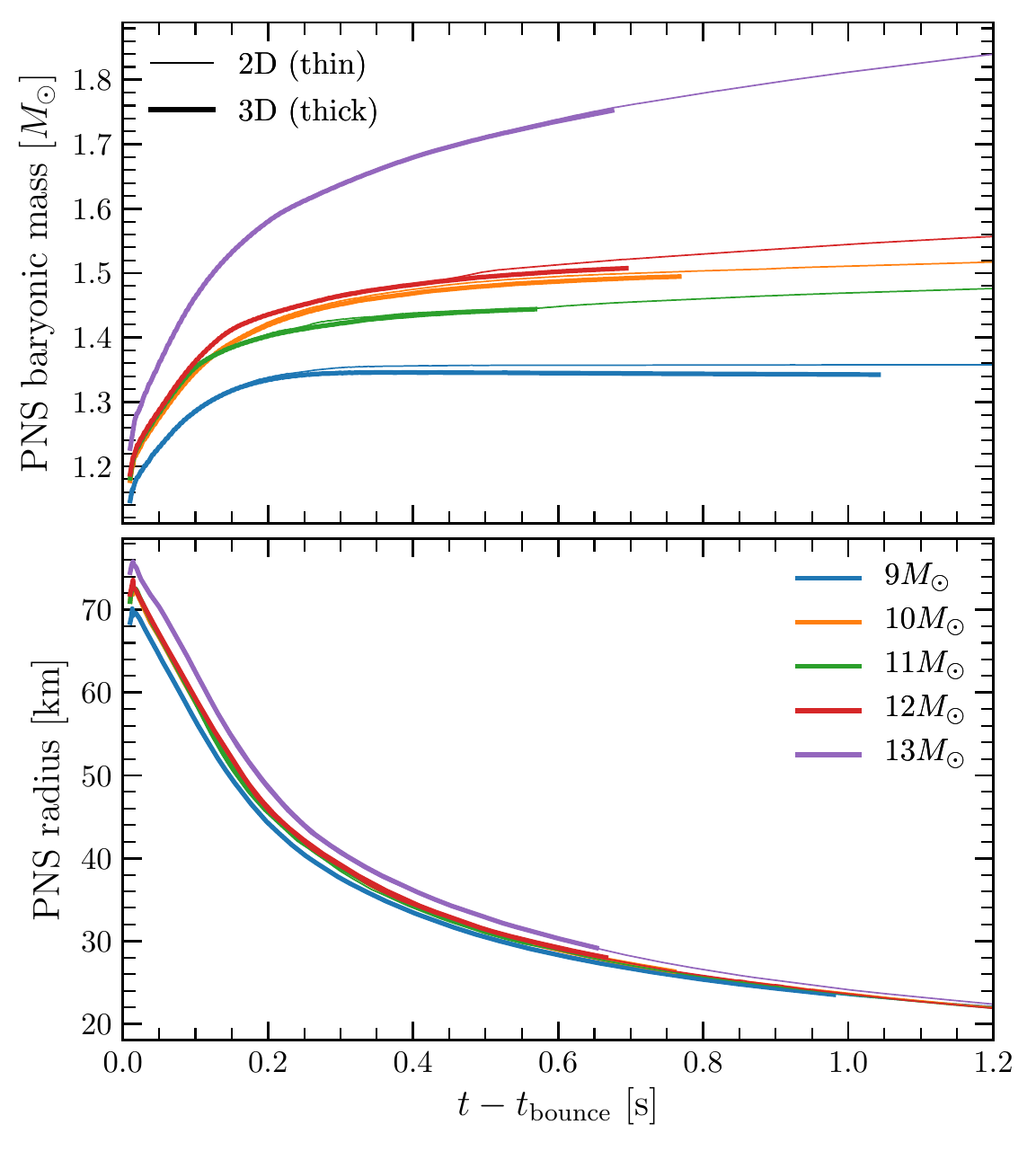}}
\caption{The corresponding proto-neutron star baryon masses (in units of M$_{\odot}$) and 
radii (in units of kilometers) versus time after bounce (in seconds). The radii
are defined as the average at a density of 10$^{11}$ g cm$^{-3}$.  The 3D models 
are the thick lines, while the 2D models are the thin lines. Note that the 
residual baryon masses for the 9-M$_{\odot}$ models have asymptoted, those
for the other models are still growing (slightly), and those for the 13-M$_{\odot}$ model
(which don't explode) are still growing quickly. The final gravitational masses (Table \ref{sn_tab})
for the exploding models range comfortably from 1.23 M$_{\odot}$ to 1.36 M$_{\odot}$.
}
\label{pns}
\end{figure*}

Figure \ref{pns} provides the evolution of both the baryon mass (top) and PNS radius (bottom)
for both the 3D (thick) and 2D (thin) models.  The final values of the residue's baryon mass
(those achieved by the end of each simulation), along with the corresponding gravitational masses,
are given in Table \ref{sn_tab}.  We see that the neutron star left behind in the 3D 9-M$_{\odot}$
explosion has a baryonic mass of 1.342 M$_{\odot}$ and a gravitational mass of 1.246 M$_{\odot}$.
These values are on the low end of the observed pulsar mass distribution and reflect the lower
accretion rates and early explosion times of this model. As suggested in Table \ref{sn_tab}, such 
low masses seem to be generic for the steeper density profiles that obtain in this low-mass 
supernova progenitor mass range.  Since the 13-M$_{\odot}$ models don't explode and the associated
progenitor density profile is the shallowest of the set (implying higher mass accretion rates), 
the corresponding 3D and 2D PNS baryon masses are large (1.714 [3D] and 1.854 [2D] M$_{\odot}$ 
at simulation end) and are still growing fast at the termination of the runs.

{\cite{bmuller_2015} observes in his long-term study of the 11.2 M$_{\odot}$ model
of \cite{woosley_heger_weaver02} that the explosion energy accumulates more unsteadily in 2D than in 3D.                
In Figure \ref{energy} in the early explosion stages, we too see that the 2D explosion energies
accumulate more unsteadily than in 3D, but at later times accumulate smoothly. Moreover,
we observe for all our models that the residual PNS masses are similar in 2D and 3D.  This is
in contradistinction to what was found in \cite{bmuller_2015} and reflects the similar 
accretion histories in 2D and 3D we witness; what differences there are seem correlated mostly with the 
(slight) differences in explosion times. An important difference with the \cite{bmuller_2015} investigation, 
other than the different code, EOS, and transport methods employed in each study, is that we include as a 
default in our microphysics suite the \cite{horowitz:17} many-body correction to the  
neutrino-nucleon scattering rates. As noted, this facilitates explosion and may mitigate 
some of the accretional differences between 2D and 3D highlighted in \cite{bmuller_2015}.
However, a full exploration of the phenomena called out in \cite{bmuller_2015} awaits a more
exhaustive study that includes longer-term simulations of a broader set of stellar 
progenitor masses.}

\begin{figure*}
\center{\includegraphics[width=0.7\textwidth]{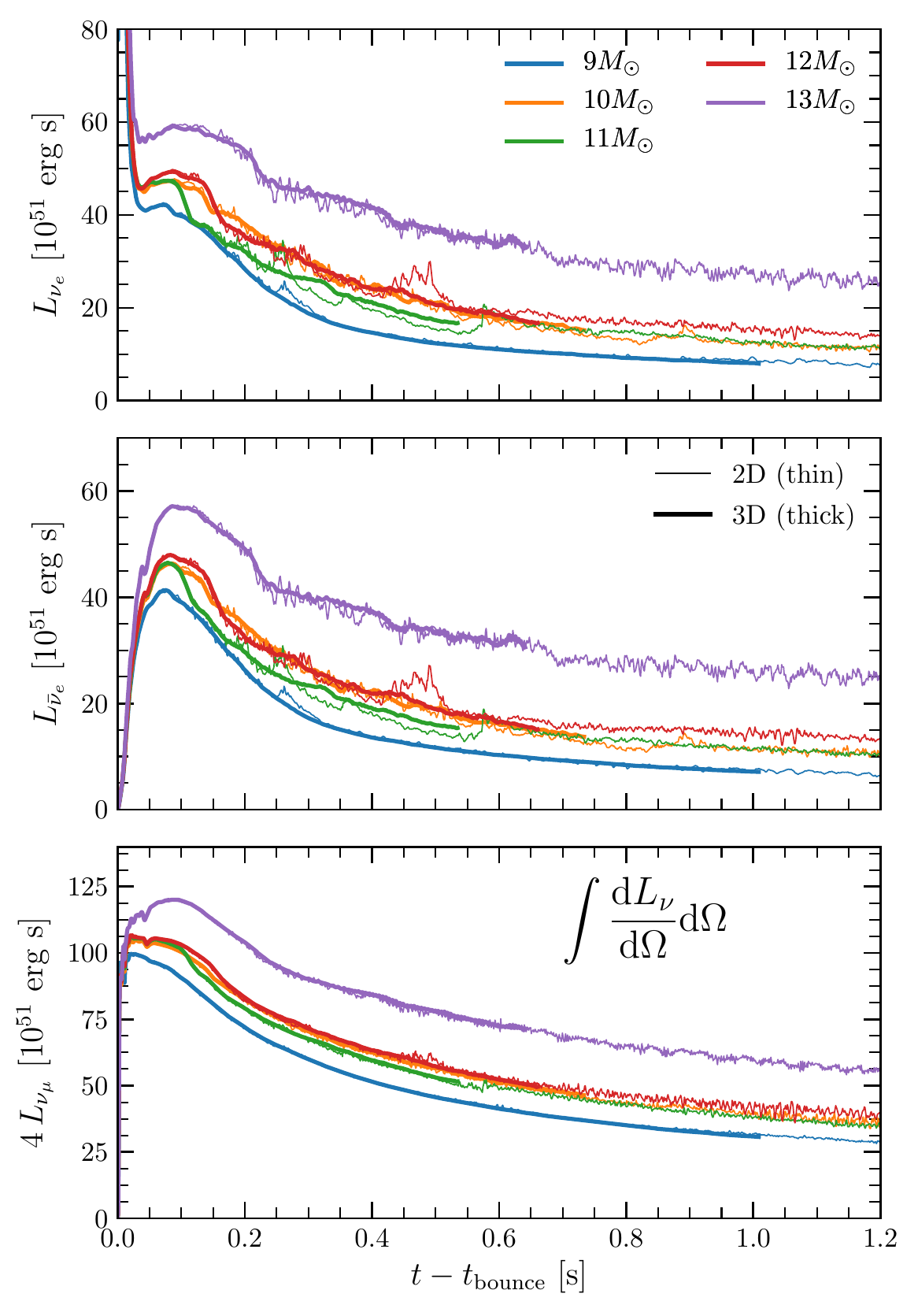}}
\caption{The evolution with time after bounce (in seconds) of the neutrino 
luminosities (in units of 10$^{51}$ erg s$^{-1}$) for the 3D (thick) and 2D (thin) 
model sets.  Shown from top to bottom are the curves for the $\nu_e$, $\bar{\nu}_e$, 
and ``$\nu_{\mu}$" species. Note that the 3D and 2D luminosity curves are solid-angle-averaged
and are quite similar.  This reflects similar core deleptonization and cooling rates and similar 
mass accretion histories (in 3D and 2D) for the quasi-spherical cores and is expected.
}
\label{luminosity}
\end{figure*}

The PNS radii are defined as the average radius at a density of 10$^{11}$ g cm$^{-3}$. As 
determined in \cite{radice:17}, due to enhanced outward transport and the alteration of inner 
entropy and Y$_e$ profiles, PNS convection increases this radius beyond what would obtain in 1D.  
The similarity of the 3D and 2D curves in Figure \ref{pns} reflects the similar luminosity and core 
energy loss rates from the interior in 3D and 2D.  This fact is reinforced by Figure \ref{luminosity},
which demonstrates that the luminosity evolutions in 2D and 3D are nearly the same.  We note
in passing that the solid-angle-integrated luminosities in 3D evolve more smoothly in time 
than in 2D, wherein axial sloshing and hydrodynamic asymmetries are more pronounced.   

\begin{figure*}
\center{\includegraphics[width=0.7\textwidth]{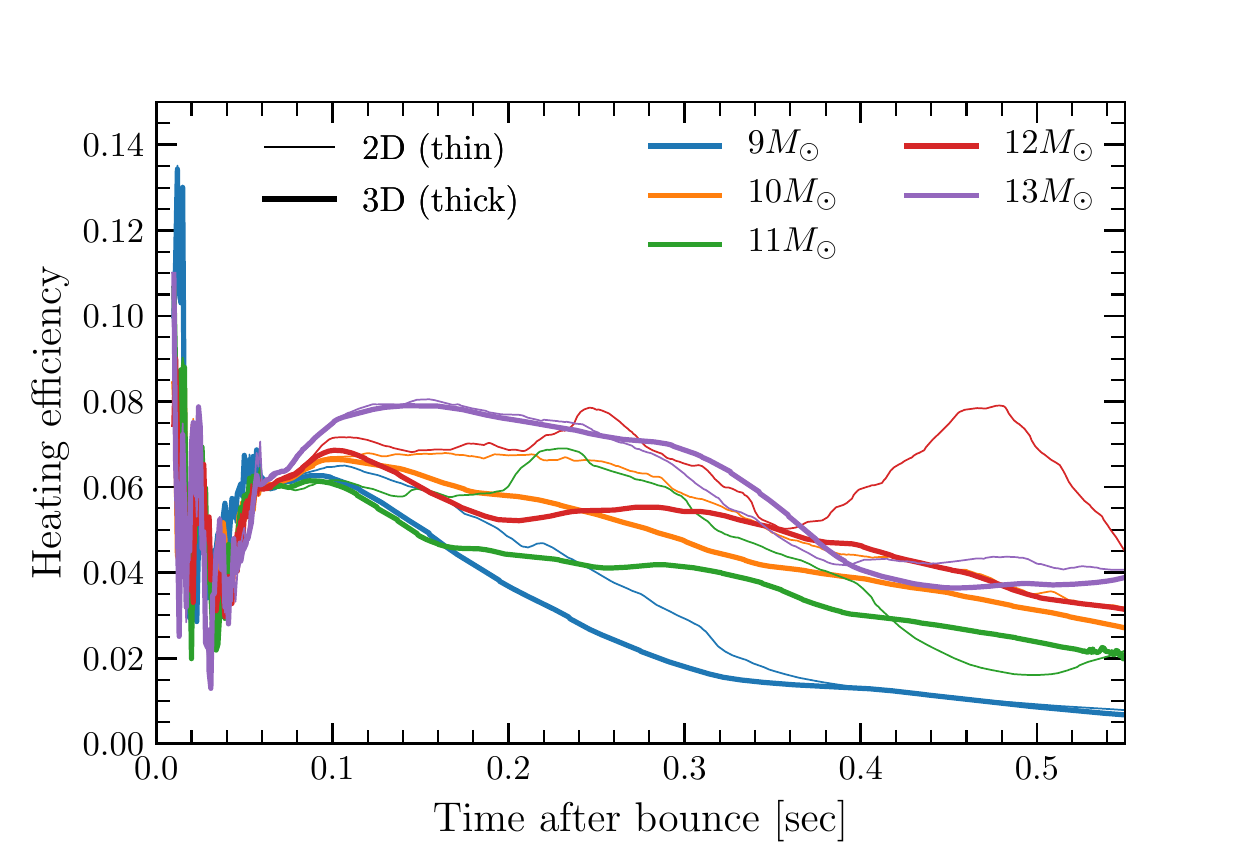}}
\caption{{The neutrino heating efficiency versus time after bounce (in seconds)
for all the 3D and 2D models simulated in this paper.  The efficiency is defined 
as the ratio of the neutrino heating rate due to $\nu_e$ and $\bar{\nu}_e$ absorption
on nucleons in the gain region and the sum of the $\nu_e$ and $\bar{\nu}_e$ luminosities.  
The efficiency ranges from a few to $\sim$8\% and is similar in 3D and 2D prior to
explosion, but differs after explosion. See text for a discussion.}
}
\label{heat}
\end{figure*}

{Figure \ref{heat} depicts the neutrino heating efficiency, defined as the ratio of the heating
rate due to $\nu_e$ and $\bar{\nu}_e$ absorption in the gain region and the sum of
the $\nu_e$ and $\bar{\nu}_e$ luminosities.  This quantity does not include the heating
due to inelastic scattering off electrons and nucleons, though this effect is included 
in the calculations and amounts to a sub-dominant $\sim$5-15\% of the total.  We see in Figure \ref{heat}
that the efficiency ranges from a few to $\sim$8\% and is similar in 3D and 2D prior to
explosion.  For the non-exploding 13-M$_{\odot}$ model, the 3D and 2D curves do not much differ
during the entire simulation, but for the exploding models, since the explosion times differ
in 3D and 2D (generally being later in 2D), the associated curves depart from one another
after the first of the models explodes.  We note that higher efficiencies do not translate into
a greater tendency to explode, as witnessed by the behavior of the 13-M$_{\odot}$ model.  We also
note that the neutrino-driven mechanism is not a ``1\% phenomenon," as frequently considered, but 
in this context is closer to a ``5$-$6\%" phenomenon.} 

\begin{figure*}
\center{\includegraphics[width=0.7\textwidth]{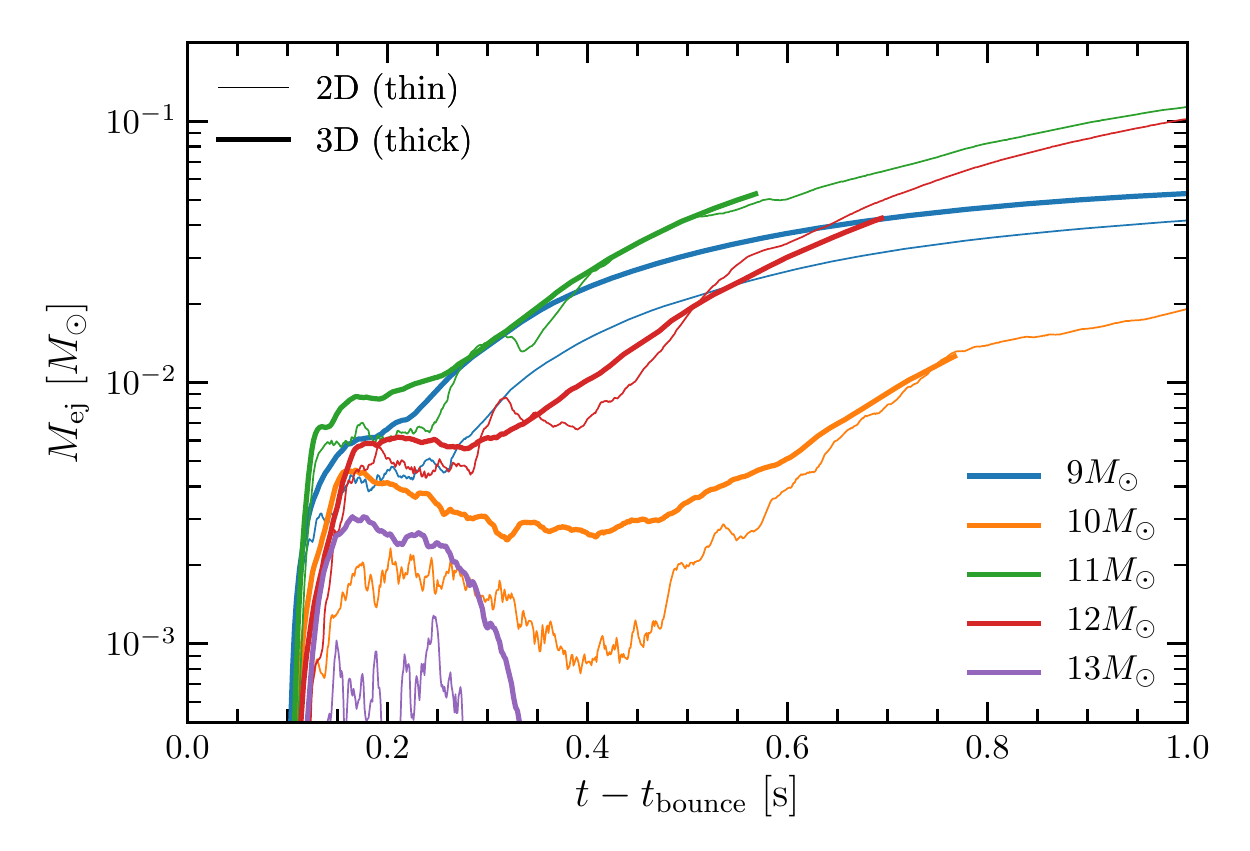}}
\caption{The ejecta mass (in units of M$_{\odot}$), defined as the mass on
positive Bernoulli trajectories, versus time after bounce (in seconds, up 
to a maximum of one second) for both the 3D (thick) and 2D (thin) models. Here, 
we are tagging only matter on our computational grid interior to the 20,000-kilometer 
radius of the initial progenitor models.  For the exploding models, it is likely
that all the matter exterior to the 20,000-kilometer outer boundary 
will in fact be ejected, but this has yet to be definitively determined. 
Note that some matter for the 13-M$_{\odot}$ models that did not eventually explode still achieved 
for a short time the Bernoulli condition we have used to tag the ejecta. 
At the end of the simulations, the ejecta mass (as defined here) for the 
exploding models ranges from $\sim$10$^{-2}$ to $\sim$10$^{-1}$ M$_{\odot}$.
}
\label{ejecta_mass}
\end{figure*}

Figure \ref{ejecta_mass} shows the evolution of the ejecta mass (in units of M$_{\odot}$)
versus time after bounce (in seconds, up to a maximum of one second) for both the 3D (thick) 
and 2D (thin) models. Only the ``qualifying" matter on our computational grid interior to the 20,000-kilometer
radius of the initial progenitor models is included in this quantity, but it is likely
that all the matter exterior to the 20,000-kilometer outer boundary for the exploding models will in fact be ejected.
Some matter for the 13-M$_{\odot}$ models that did not eventually explode still achieved
for a short time the Bernoulli condition we have used to identify the ejecta.
At the end of the simulations, the ejecta mass (as defined here) for the
exploding models ranges from $\sim$10$^{-2}$ to $\sim$10$^{-1}$ M$_{\odot}$.

\begin{figure*}
\center{\includegraphics[width=0.7\textwidth]{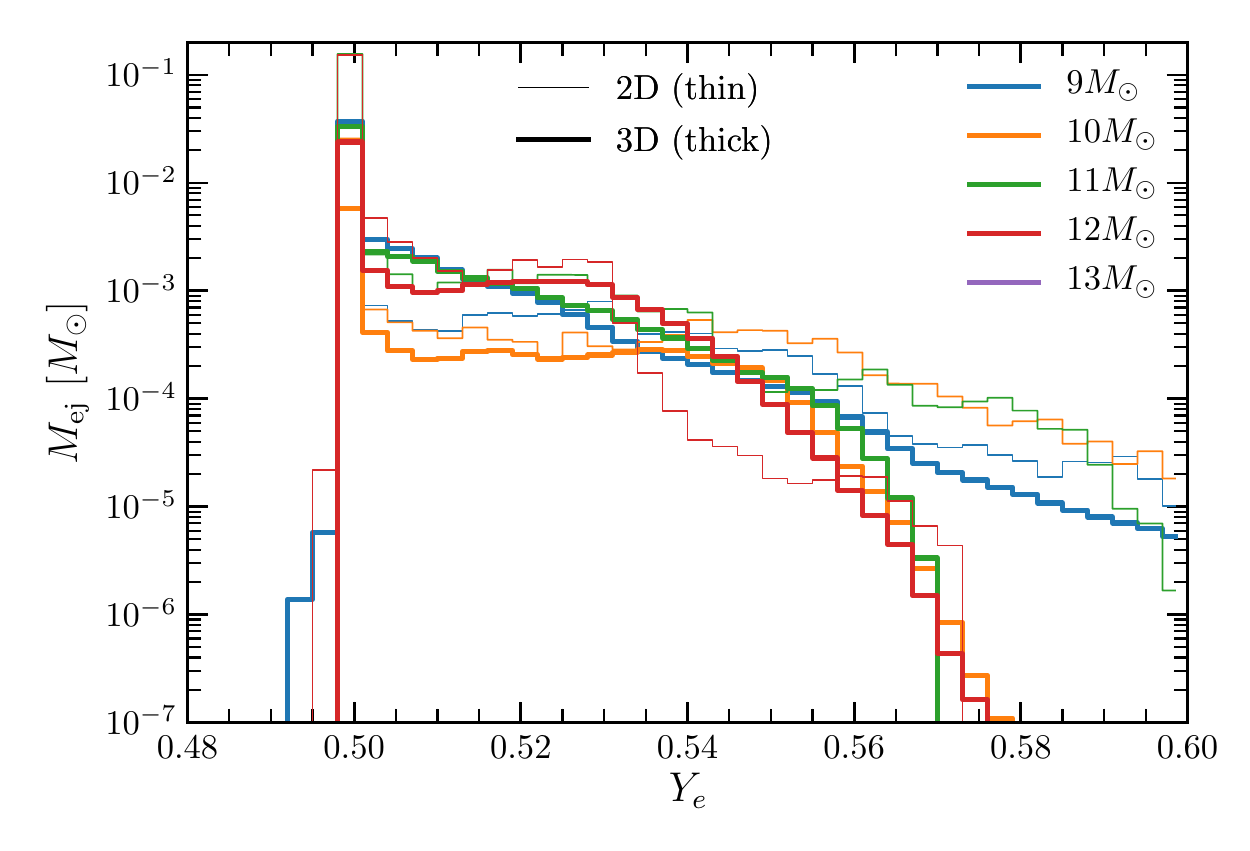}}
\caption{Histograms of the final (end of simulation, see Table \ref{sn_tab}) 
Y$_e$ distributions for the 3D (thick) and 2D (thin) models calculated.
This is a true histogram for which the total ejecta mass in a particular 
Y$_e$ bin is the mass given on the ordinate (y-axis).  M$_{\rm ej}$
on the ordinate is not a distribution function in Y$_e$ (i.e., not 
$\frac{dM}{dY_e}$).  We note that the total ejecta mass 
with Y$_e$ = 0.5 for the 3D models ranges between $\sim$5$\times$ and $\sim$40$\times$10$^{-2}$ 
M$_{\odot}$ and provides a scale for the $^{56}$Ni mass ejected.  However, we 
also note that we did not perform true nucleosynthetic calculations for these runs
and are leaving such studies to an upcoming new generation of 3D calculations.
}
\label{histograms}
\end{figure*}

Figure \ref{histograms} renders histograms of the final Y$_e$ distributions for the 3D (thick) 
and 2D (thin) models calculated. This is a true histogram for which the total ejecta mass 
in a particular Y$_e$ bin is the mass given on the y-axis.  Importantly, M$_{\rm ej}$ on 
the ordinate is not a distribution function in Y$_e$ (i.e., not $\frac{dM}{dY_e}$)
and the actual sum of the histogram values given here is the last total ejecta 
mass plotted in Figure \ref{ejecta_mass} and/or given at the final time listed in 
Table \ref{sn_tab}.  We find that the total ejecta mass with Y$_e$ = 0.5 for 
the 3D models ranges between $\sim$5$\times$ and $\sim$40$\times$10$^{-2}$ M$_{\odot}$ and 
provides a scale for the $^{56}$Ni mass ejected.  However, we did not perform 
nucleosynthetic calculations for these runs and are leaving such studies 
to an upcoming new generation of 3D calculations. As Figure \ref{histograms}
shows, the ejecta for the 3D and 2D runs are not vastly different and are generally
proton rich, extending up beyond Y$_e$ = 0.56, with most of the ejecta near 0.5.
The ejecta Y$_e$s are a consequence of the competition between $\nu_e$ and $\bar{\nu}_e$
absorption and, hence, quite dependent on the fidelity with which the neutrino-matter
coupling is handled. This fact should be borne in mind and puts a premium
on accurate neutrino transport.

\section{Explosion Morphologies}
\label{debris}

\begin{figure*}
\includegraphics[width=0.45\textwidth]{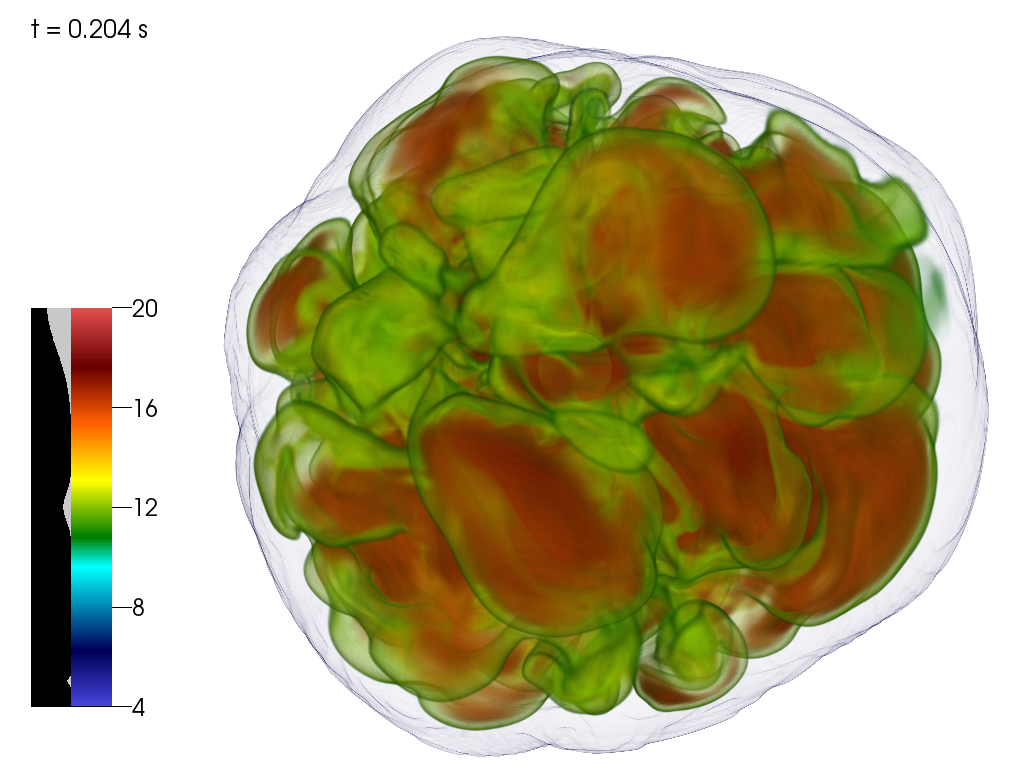}
\includegraphics[width=0.45\textwidth]{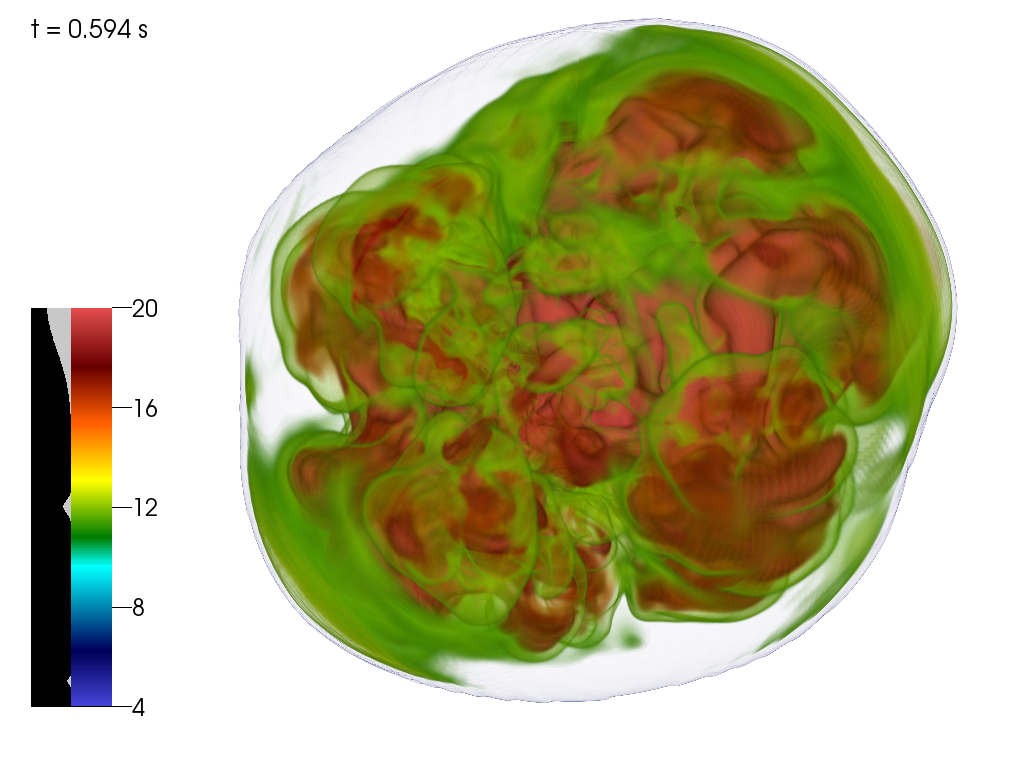}
\caption{Two representative stills during the post-bounce 3D evolution of the exploding 
9-M$_{\odot}$ model. Time proceeds from left to right and the spatial scale expands
as a function of time.  The outer blue shroud is the shock wave. The representation 
is a volume rendering of the entropy at the post-bounce time given in each 
top-left corner (in seconds) and the associated color map given in the bottom-left corner.
The entropy units are per baryon per Boltzmann's constant.  High entropies in the shocked mantle are
more conducive to explosion, but entropy alone does not determine a predilection towards explosion.
The physical scales {(approximate diameter of the shock)} are 400 km (left) and 6000 km (right).
Note that, as with the following figures, the last time depicted here is not the last 
time of the simulation (see Table \ref{sn_tab}). See the text for a discussion of this and related plots.
}
\label{3D_9}
\end{figure*}

\begin{figure*}
\includegraphics[width=0.45\textwidth]{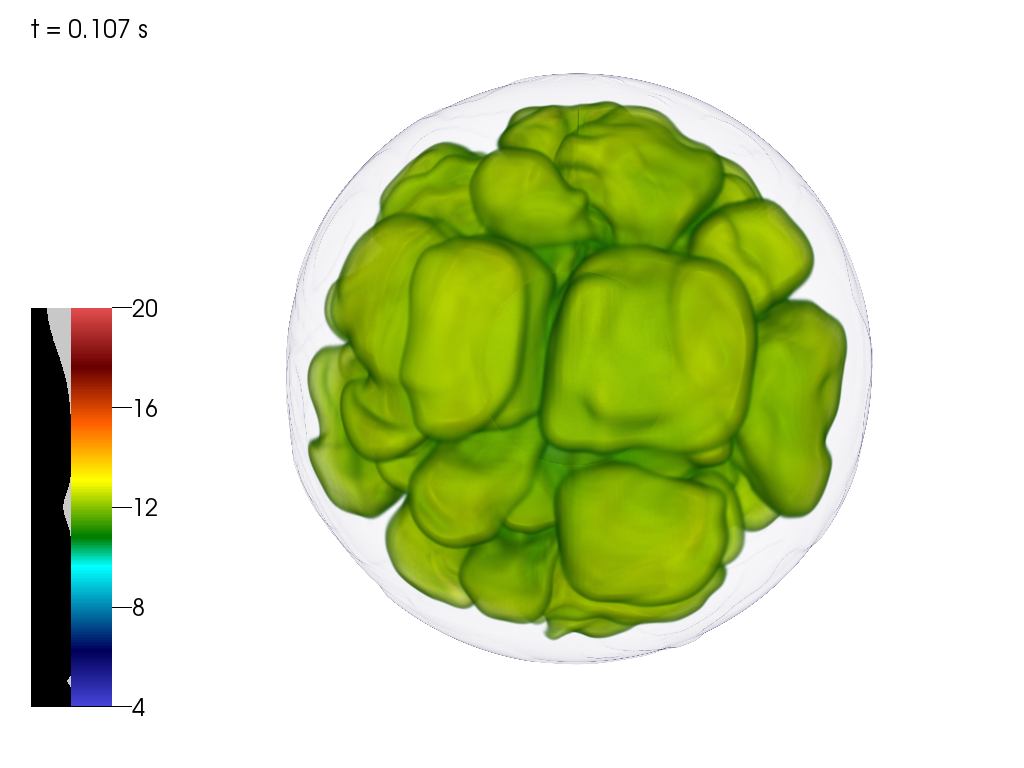}
\includegraphics[width=0.45\textwidth]{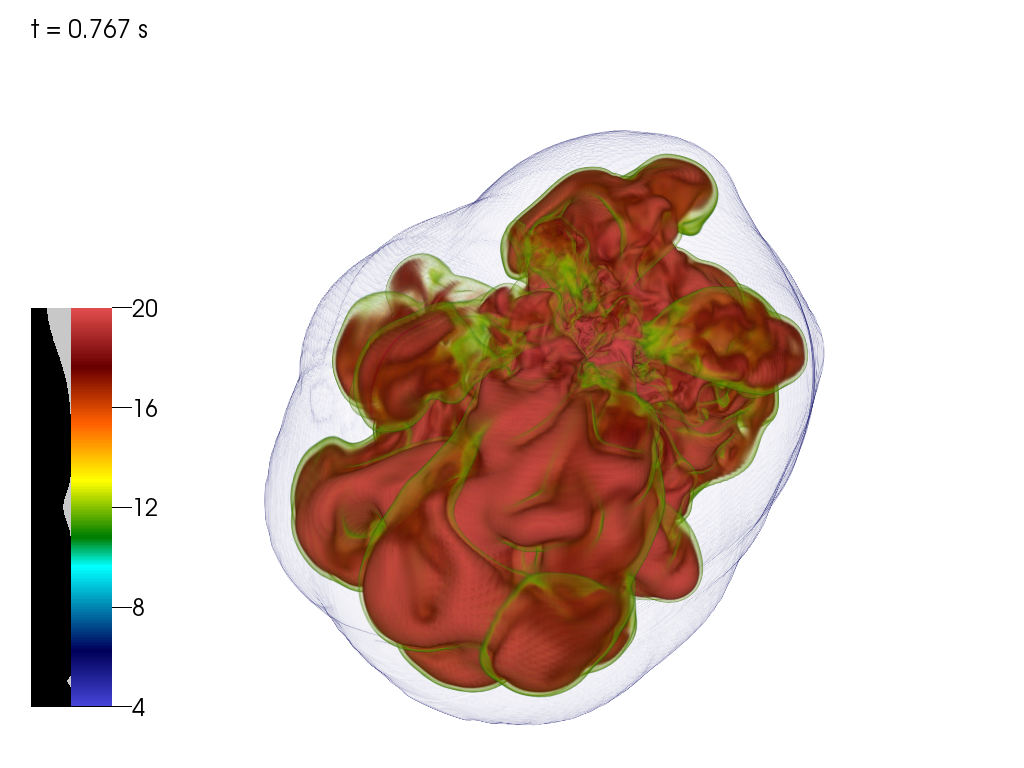}
\caption{Same as Figure \ref{3D_9}, but for the
10-M$_{\odot}$ model.  Note that the entropy scales are the same as in Figure \ref{3D_9}, but that 
snapshot times are different. The physical scales  {(approximate diameter of the shock)} are 200 km (left) and 2700 km (right). 
}
\label{3D_10}
\end{figure*}

\begin{figure*}
\includegraphics[width=0.45\textwidth]{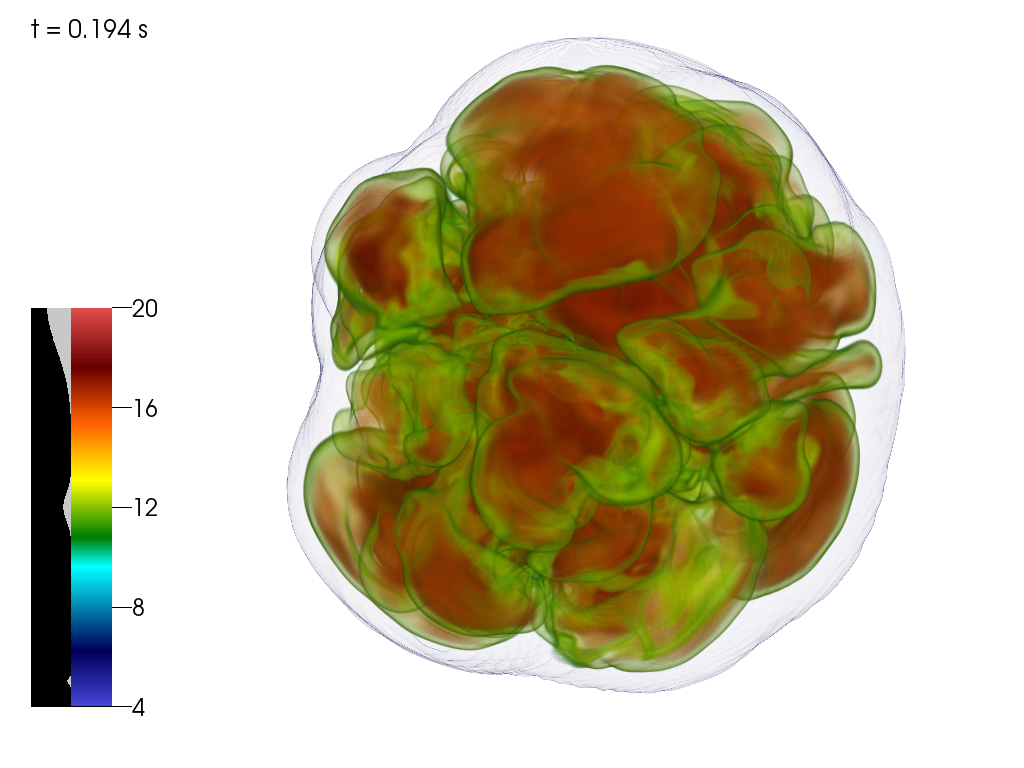}
\includegraphics[width=0.45\textwidth]{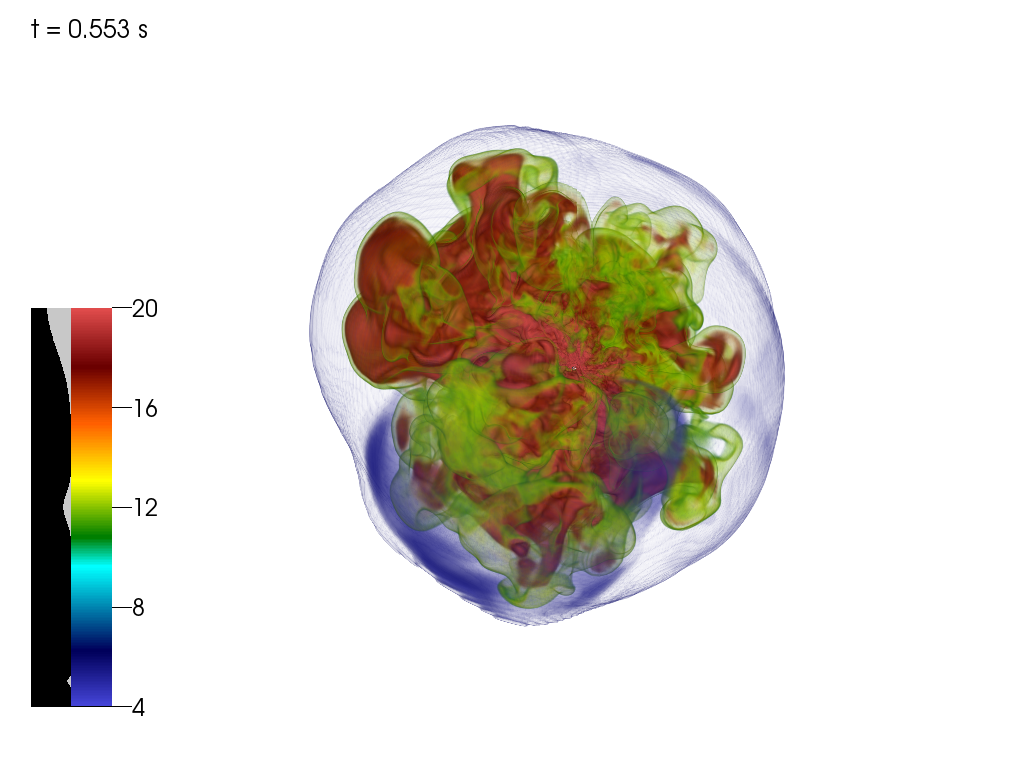}
\caption{Same as Figure \ref{3D_9}, but for the
11-M$_{\odot}$ model.  Note that the entropy scales are the same as 
in Figure \ref{3D_9}, but that snapshot times are different. 
The physical scales  {(approximate diameter of the shock)} are 420 km (left) and 4000 km (right).
}
\label{3D_11}
\end{figure*}

\begin{figure*}
\includegraphics[width=0.45\textwidth]{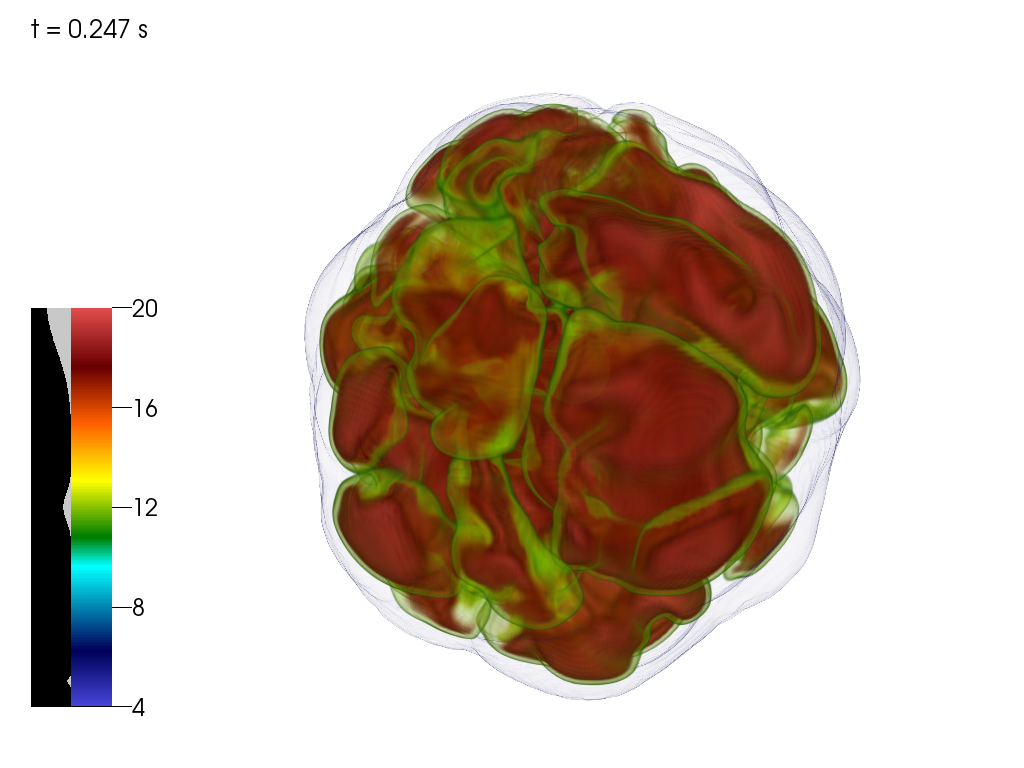}
\includegraphics[width=0.45\textwidth]{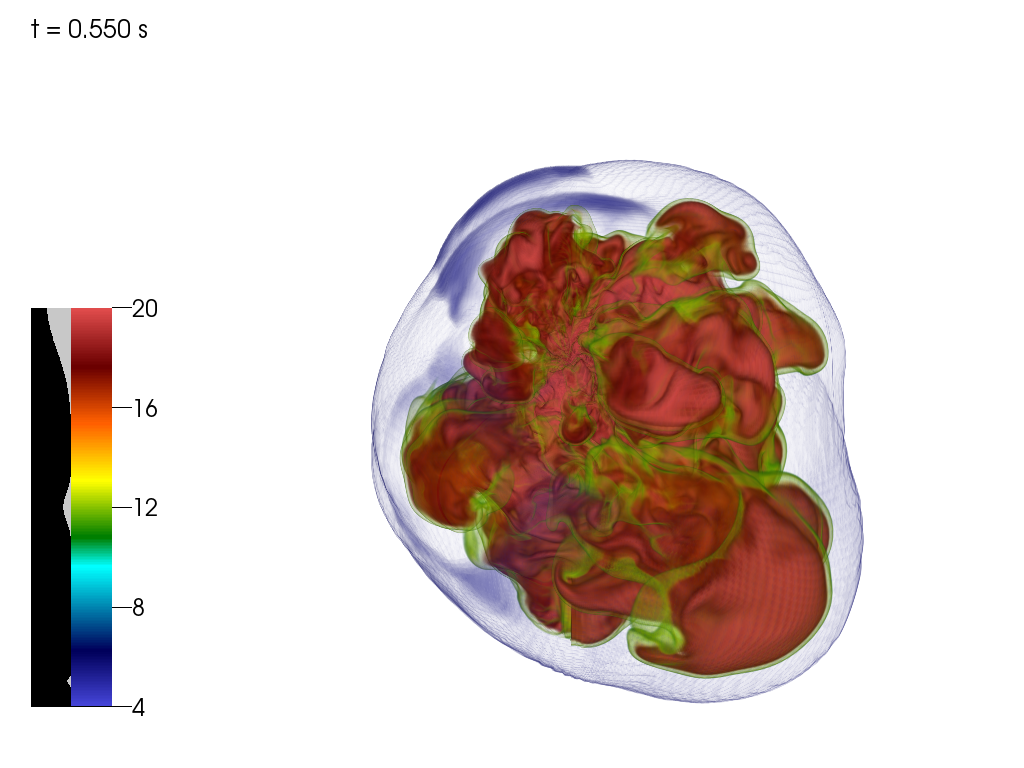}
\caption{Same as Figure \ref{3D_9}, but for the
12-M$_{\odot}$ model.  Note that the entropy scales are the same as in Figure \ref{3D_9}, 
but that snapshot times are different.  The physical scales  {(approximate diameter of the shock)} 
are 436 km (left) and 2500 km (right).
}
\label{3D_12}
\end{figure*}

\begin{figure*}
\includegraphics[width=0.45\textwidth]{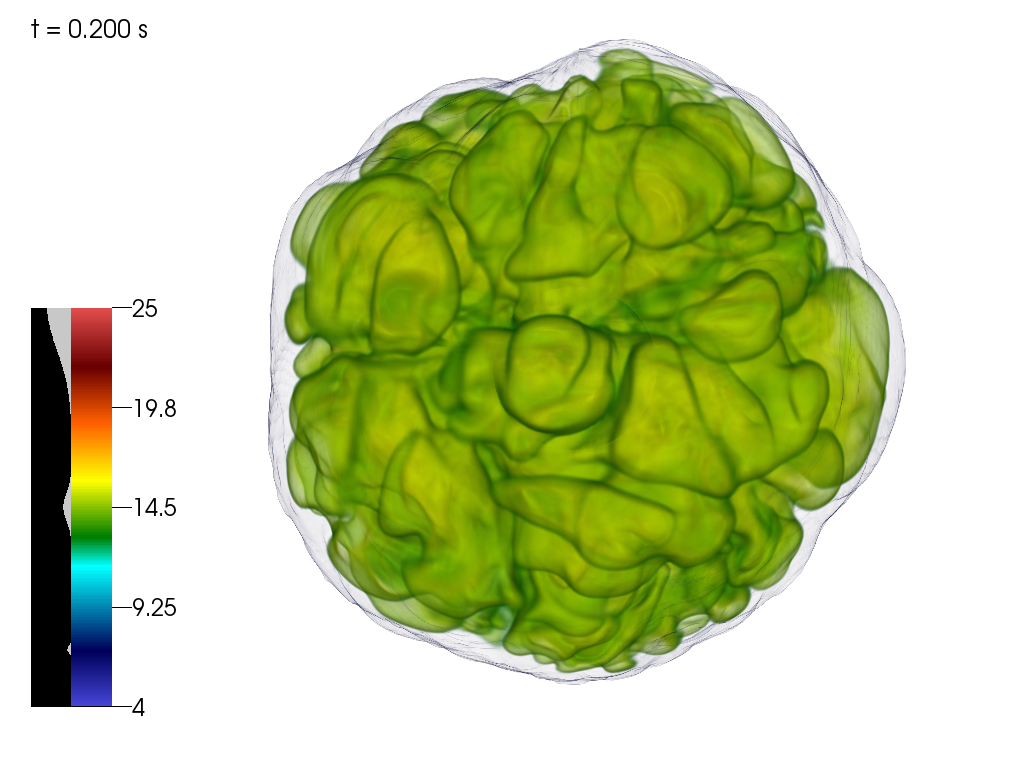}
\includegraphics[width=0.45\textwidth]{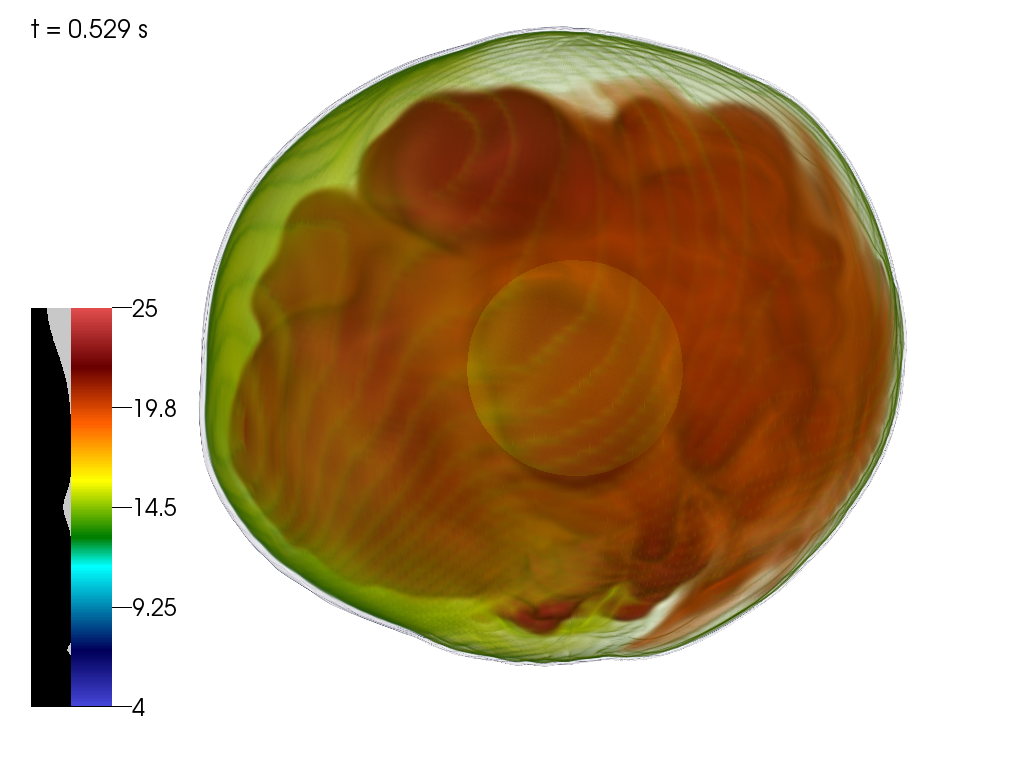}
\caption{Same as Figure \ref{3D_9}, but for the
13-M$_{\odot}$ model.  Note that the entropy scales are the same as in Figure \ref{3D_9}, but that           
snapshot times are different.  By that the end of the simulation this model had not exploded and the shock
radius had shrunk significantly. The circles dimly seen in the centers of the two rightmost figures
trace the PNS "surface," defined as the surface where $\rho = 10^{11}$ g cm$^{-3}$. As indicated in
Table \ref{sn_tab2}, at the last time depicted here the PNS radius was $\sim$30 kilometers.
The physical scales  {(approximate diameter of the shock)} are 200 km (left) and 100 km (right).
}
\label{3D_13}
\end{figure*}

Figures \ref{3D_9} through \ref{3D_12} depict snapshots of volume 
renderings of the entropy distributions for the 9-M$_{\odot}$, 10-M$_{\odot}$,
11-M$_{\odot}$, and 12-M$_{\odot}$ models interior to the shock wave. 
This post-shock mantle material 1) first participates in the turbulent 
convection exterior to the PNS core and interior to the stalled shock 
prior to explosion (left slide) and then 2) constitutes, along with 
the outer matter encompassed by the expanding shock, the neutrino-driven 
bubbles and mushroom clouds of the subsequent explosion (right slide). 
The outer blue shroud is the shock wave. Recall that the 13-M$_{\odot}$ model 
(Figure \ref{3D_13}) does not explode and that both its shock wave and 
PNS core eventually shrink as the core deleptonizes and cools.

\begin{figure*}
\includegraphics[width=0.7\textwidth]{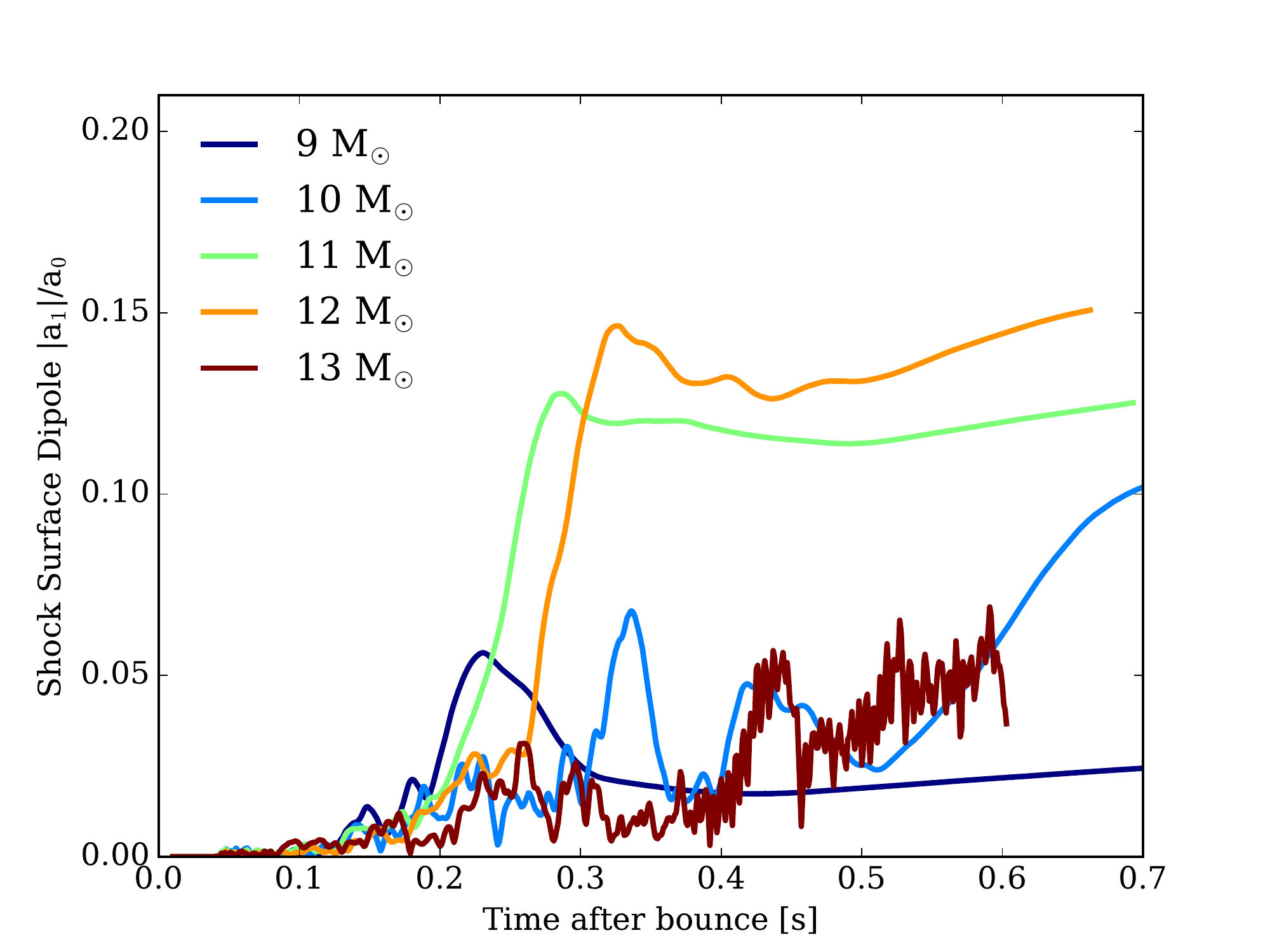}
\caption{The amplitude of the dipole component of the shock surface, normalized
by the corresponding monopole term, as a function of the time after bounce (in seconds).
The algorithm for calculating this quantity is taken from \protect\cite{burrows2012}.
The five 3D models of this plot are indicated by different colors.  Note
that the ramp up of the dipolar component (when it occurs) roughly
coincides with the onset of explosion (Table \ref{sn_tab}).
See the text for a discussion.
}
\label{dipole}
\end{figure*}

Prior to explosion, the characteristic physical scale of the neutrino-driven 
turbules behind the shock is set by the size of the gain region.  This is tens
of kilometers to $\sim$100 kilometers and generally represents less than or
equal to 50\% of the shock radius (to the center).  Hence, the angle subtended
by the turbules is tens of degrees, which translates into modes of $\ell$ = 5$-$10.
To be sure, this is a crude estimate, but serves as a useful mnemonic.\footnote{The
growth rates in the earliest phases of instability development will depend 
upon the initial perturbation spectrum.}  However, as the explosion commences, the dominant
harmonic components of the shock position itself are predominantly 
the monopole (clearly) and the dipole.  In fact, most exploding models have a
growing dipole \citep{dolence:13}, whose growth seems in phase with that
of the monopole (whose growth itself constitutes a zeroth-order marker of
explosion).  For these non-rotating models, the direction chosen by the dipole
(a vector) seems arbitrary and is not correlated with the coordinate axes (fortunately),
but is correlated with the kick direction (Radice et al. 2019, in preparation). This
emphasizes the conclusion that the explosions of such stochastic and chaotic models 
without physically defined directions would in reality span a uniform spectrum 
of directions. Figure \ref{dipole} depicts the monopole-normalized
dipole magnitude versus time after bounce for the five 3D models of this paper.
The harmonic decomposition employed to generate this figure is that found in 
\cite{burrows2012} and \cite{vartanyan2019}.  The explosions of our 
11-M$_{\odot}$ and 12-M$_{\odot}$ models are particularly asymmetrical
and dipolar, with the dipole component of the shock surface of the 10-M$_{\odot}$ model 
growing fast at the end of that simulation, while by that metric the 3D 9-M$_{\odot}$ 
model is much more spherical.  That the latter is more spherical makes sense, 
since this model explodes almost immediately and a dipole mode doesn't 
seem to have time to grow. 

\begin{figure*}
\includegraphics[width=0.45\textwidth]{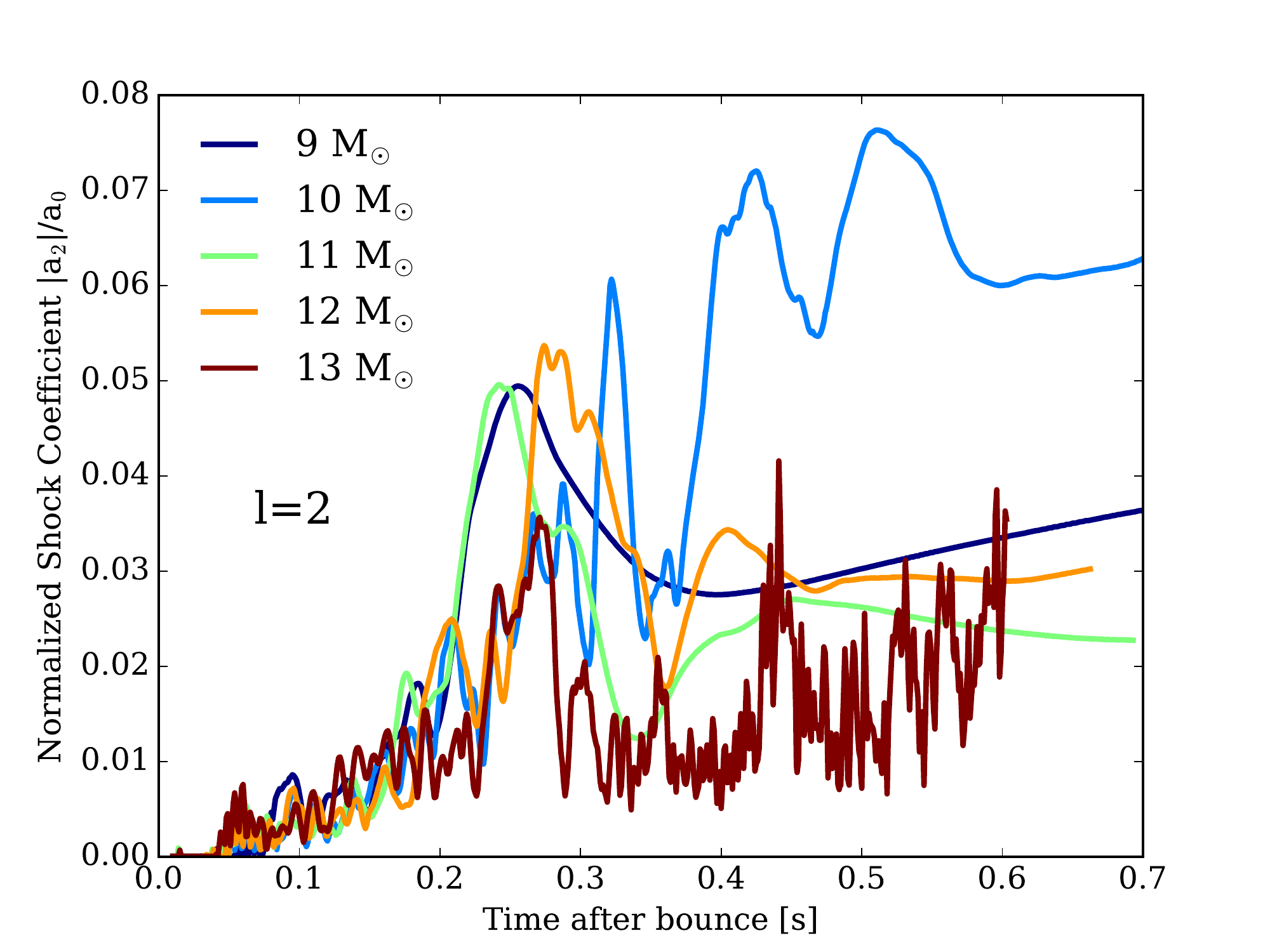}
\includegraphics[width=0.45\textwidth]{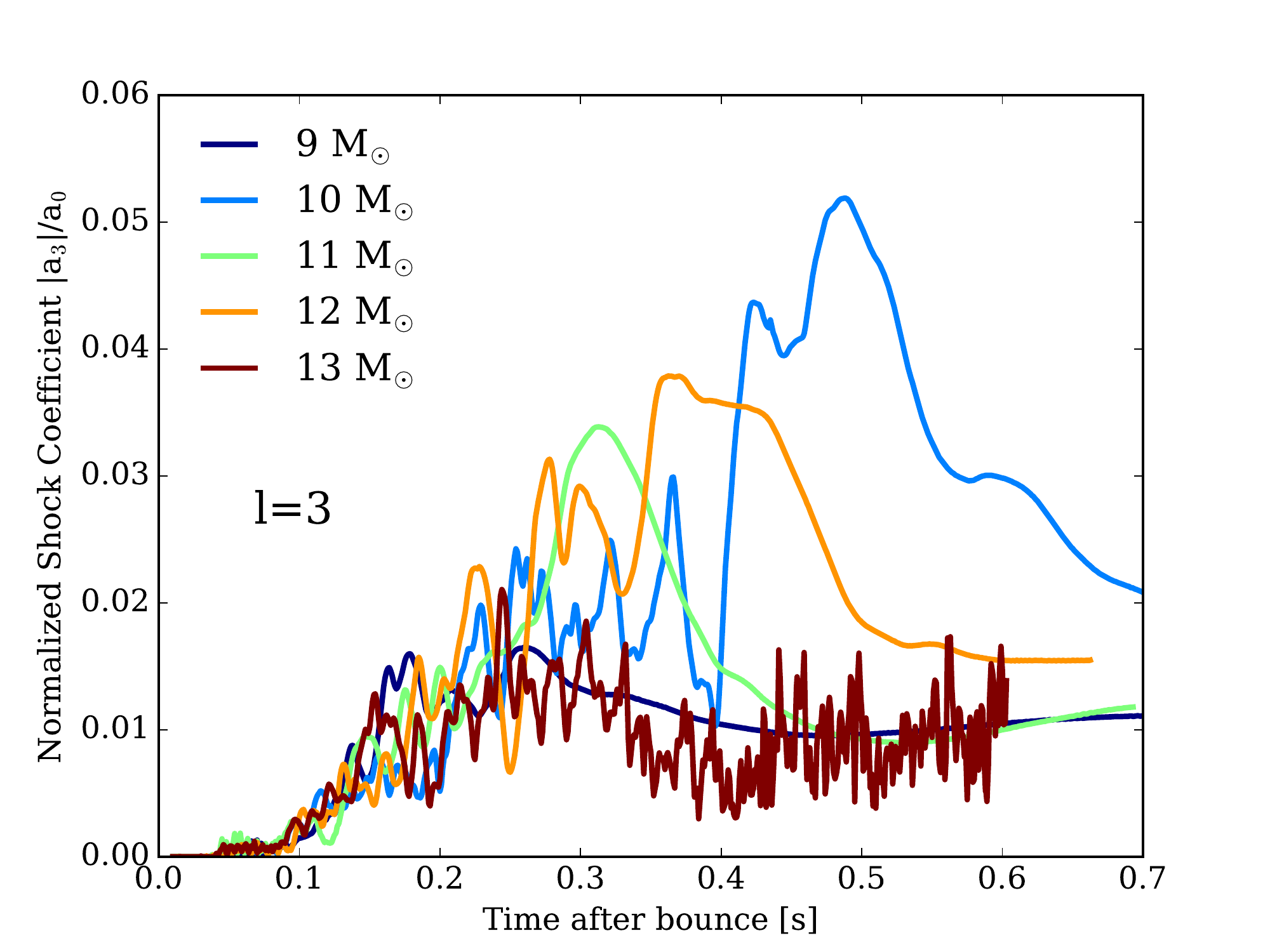}
\includegraphics[width=0.45\textwidth]{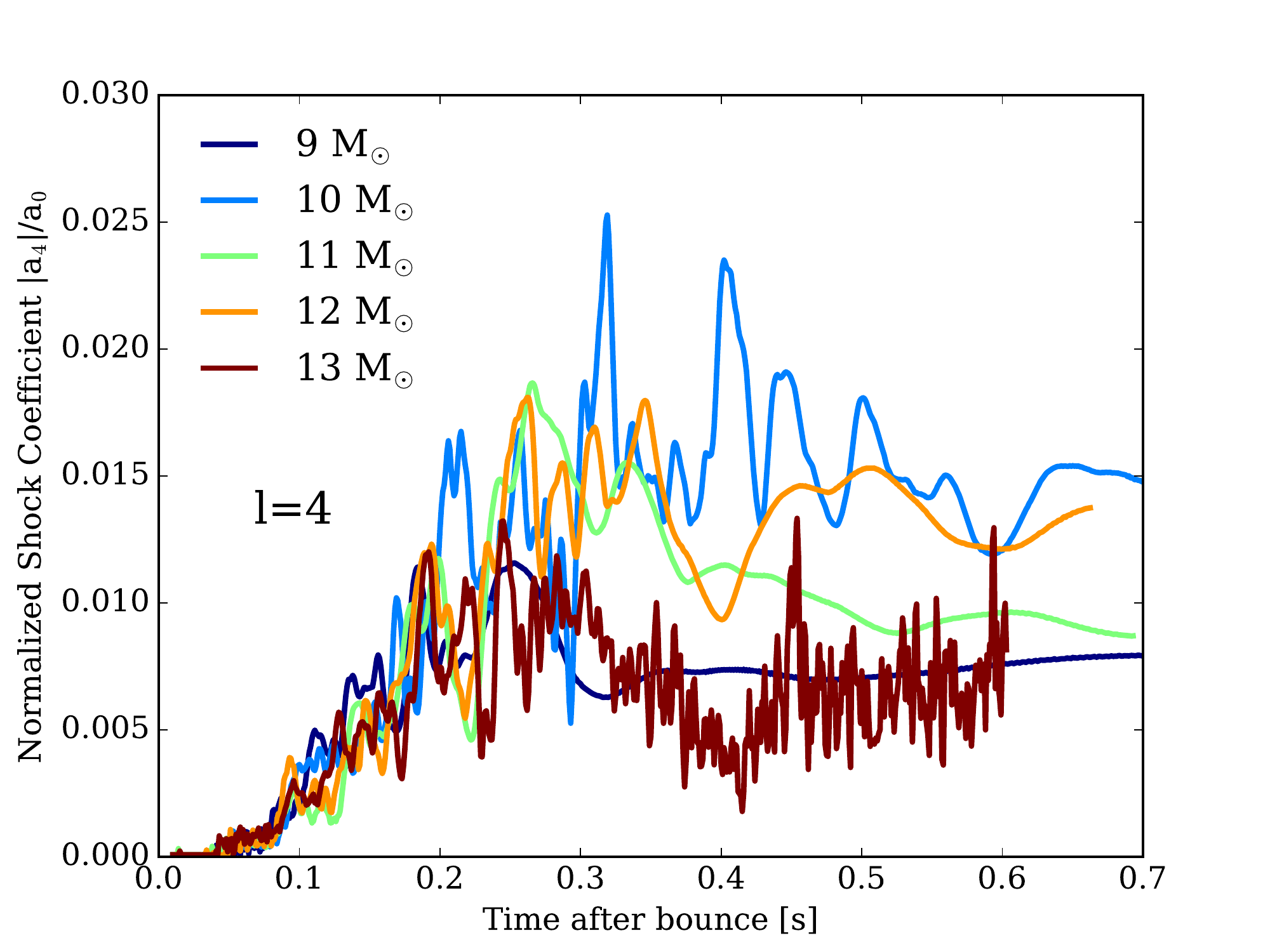}
\includegraphics[width=0.45\textwidth]{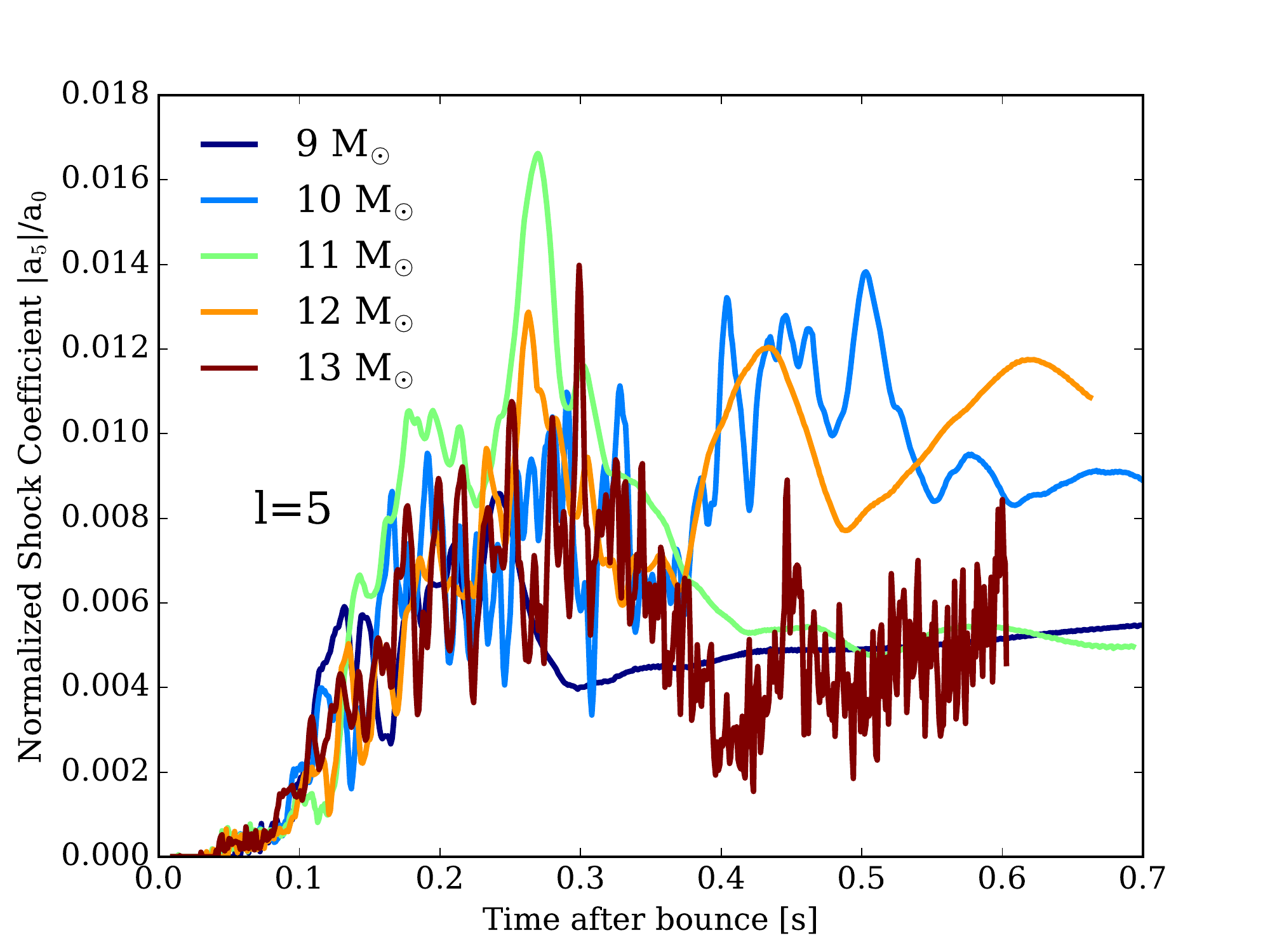}
\caption{Similar to Figure \ref{dipole}, but for the $\ell = 2,3,4,5$
harmonic coefficients of the shock surface (normalized to the monopole term)
versus time since bounce (in seconds).  $\ell = 2$ is in the top left, $\ell = 3$ is in the top
right, $\ell = 4$ is in the bottom left, and $\ell = 5$ is in the bottom right.
As in \protect\cite{burrows2012}, for the 3D models and the various $\ell$s
the $m$ subcomponent terms are added in quadrature.  See the text for a discussion.
}
\label{multi}
\end{figure*}

\begin{figure*}
\includegraphics[width=0.60\textwidth]{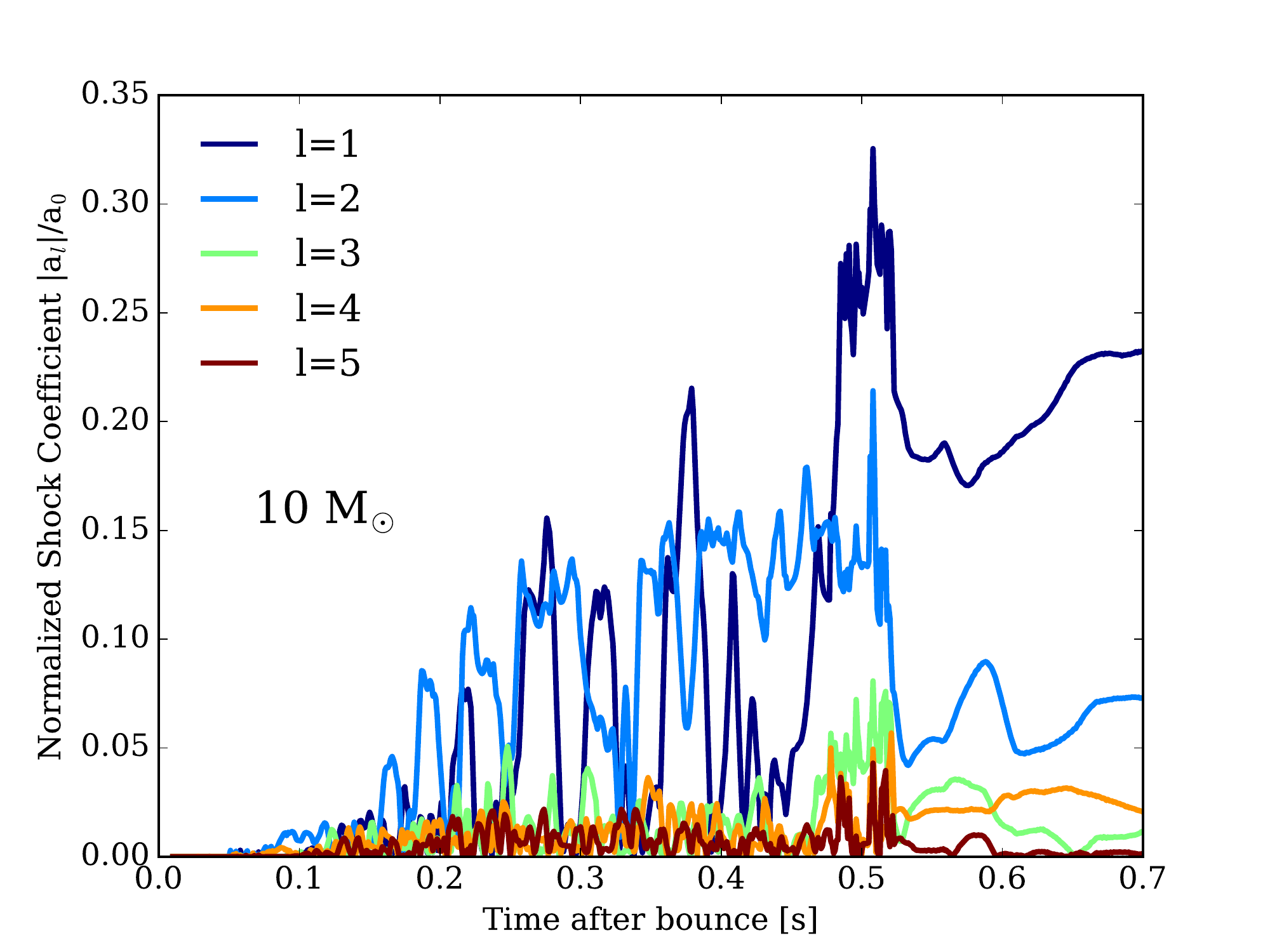}
\caption{Similar to the panels in Figure \ref{multi}, but for the $\ell = 1,2,3,4,5$
harmonic coefficients of the shock surface (normalized to the monopole term) of the 2D
10-M$_{\odot}$ model.  See the text for a discussion.
}
\label{multi_2D}
\end{figure*}

{Figure \ref{multi} renders the corresponding evolution for the normalized higher-order
spherical harmonics ($\ell = 2,3,4,5$) of the deformation of the shock surface in 3D.
For the 3D model results in both Figures \ref{dipole} and \ref{multi}, we have lumped
together in quadrature the azimuthal $m$s in an effort to avoid the significant clutter
that would attend their separate inclusion.  However, there is much information in this
more refined azimuthal decomposition and we will explore in the future ways to represent it.

We observe that for a given progenitor model the magnitude of the higher-order harmonic coefficients
generally decreases as $\ell$ increases and that for the non-exploding 13-M$_{\odot}$ model
all $\ell$s below six remain of rather low relative amplitude.  As both Figures \ref{dipole}
and \ref{multi} demonstrate, all the non-monopolar components start small and grow with
time as turbulent convection builds.  The speed with which this occurs reflects in part
the character and magnitude of the seed perturbations, which for all our models are
quite small.  After explosion onset, for all models the higher-order normalized
$\ell$ coefficients grow and then saturate, or grow, peak, slightly diminish, and 
then saturate (or assume a more secular, long-term drift), with the dipole and 
quadrupole terms usually predominating in relative magnitude. The time at which 
the normalized amplitudes of the various harmonics peaks correlates with the 
explosion time, with the coefficients for the 9-M$_{\odot}$ model evolving most quickly
through its stages.  For this most spherical of our 3D models, the amplitudes of the higher-harmonics
remain rather small. For the more-delayed 10-M$_{\odot}$ model, the dipole term remains small
until explosion is fully underway, and then strengthens (as it does for those models that
explode earlier). Before this phase, its normalized shock quadrupole is roughly comparable to its
corresponding dipole. Curiously, only the dipole component is of real significance for the
11-M$_{\odot}$ model; all its other modes peak early and then assume rather small values.
For all models, the temporal fluctuations of all components are significant before
explosion, but damp out after explosion as the ejecta evolve into a more coasting, nearly
homologous, phase.  For the non-exploding 13-M$_{\odot}$ model, fluctuations remain
rapid on the roughly $\sim$10-millisecond timescales of the turbulence in the gain region.
For comparison, we portray in Figure \ref{multi_2D} the evolution of the normalized $\ell = 1,2,3,4,5$ 
shock harmonics for the 10-M$_{\odot}$ model in 2D. This choice of progenitor is meant merely to 
be representative.  We notice immediately the larger amplitude of the temporal fluctuations 
in 2D until the simulation is well into explosion, the clear hierarchy in harmonic order with $\ell$, 
and the larger asymptotic values of the dipole and quadrupole coefficients in 2D than in 3D.
The latter are clearly artifacts of the 2D constraint.}

We emphasize that all explosions are dominated interior 
to the shock by bubble structures in mass and entropy, not surprisingly very much 
like mushroom clouds, with higher harmonic order and angular scales of 
$\sim$30$-$60 degrees.  At late times, but before instabilities in
the outer reaches of the massive star can come into play \citep{utrobin_2018}, 
the bubble structures freeze into a quasi-homologous expansion with frozen 
angular scales, {reflecting the approximate freezing of the associated shock 
surface structure}. However, we note that the influence of computational resolution 
on the morphology and angular scales of the debris has yet to be determined.

As seen in our 3D 16-M$_{\odot}$ explosion model \citep{vartanyan2019}, 
two lobes sometimes characterize the debris field, with a ``wasp-like" waist separating
the two exploding lobes of different sizes {(see also \cite{bmuller_2015}, his Figure 6)}.  
We see something like this for the 10-M$_{\odot}$ and 12-M$_{\odot}$ 3D models, but not for the 
9-M$_{\odot}$ and 11-M$_{\odot}$ 3D models and it is curious (perhaps coincidental?) 
that those former models explode the latest.  For a time during the early explosion, 
this waist, when it appears, is the region where matter is still accreting onto the PNS core in an annular 
region.  This continued accretion helps maintain the driving $\nu_e$ and $\bar{\nu}_e$ neutrino 
luminosities, which is emitted more isotropically, despite the directed character of 
wasp-waist accretion. Such annular accretion seems to funnel matter away from the exploding
lobes, thereby diminishing the tamp they experience and, perhaps, facilitating explosion.
In some sense, this partial funnelling of matter from antipodal regions into which matter
is exploding to an annular region that, at early times, is not exploding is akin to a quadrupolar 
($\ell = 2$) symmetry breaking in the velocity field and may be a mode found by the outer flow to 
encourage explosion.  Numerous questions emerge, among which are: 1) Is it the case that some 
models employ such a quadrupolar instability to facilitate explosion?  and 2) Given 
a progenitor, how predictable is the character of the debris field and what is 
the distribution of explosion morphologies for a given star?  We don't yet know the answers 
to these questions, but these are worthy future lines of investigation.

The debris fields and morphologies that we have described and that are represented in 
Figures \ref{3D_9} through \ref{3D_12} are but the initial conditions for the $^{56}$Ni-bubble,
Rayleigh-Taylor, and Richtmyer-Meshkov instability phases and reverse shocks 
experienced by the ejecta \citep{fryxell1991,2003A&A...408..621K,utrobin_2018} at later times 
as the blast traverses the outer star.  We also note that, aside from the 9-M$_{\odot}$ 
model, our exploding 3D models have not yet asymptoted to their final energies and PNS configurations, 
though they have achieved positive total explosion energies. Finally, for the non-exploding 13-M$_{\odot}$ model,
there is some indication at late times of the development of a spiral ``SASI" mode \citep{blondin_shaw,BlMe07,rantsiou}
and whether this will transition into an explosion at very late times \citep{takiwaki2016}, 
though unlikely, has yet to be determined.  Other than this late-time manifestation 
in our 3D 13-M$_{\odot}$ run, we don't witness the classical SASI \citep{blondin2003} 
in any of our simulations.  This is in keeping with the conclusion by \cite{burrows2012}
that the SASI is more likely to be in evidence for non-exploding models and that 
in exploding models neutrino-driven convection generally overwhelms it. However,
the final word on the appearance and importance of the SASI has yet to be written.

\section{Conclusions}
\label{conclusions}

Our new 3D results, and those 1D, 2D, and 3D results of others 
\citep{kitaura,burrows2007_onemg,fischer:10,muller:2012,melson:15a,radice:17}
for this lower-mass range, now suggest that these stars can easily explode by the delayed neutrino
mechanism with explosion energies not far from what is observationally expected 
\citep{morozova_energy}.  Moreover, with the results articulated here and elsewhere 
in the recent literature (e.g., \cite{burrows:18,vartanyan2019}), 
we can conclude that various features and/or processes disproportionately 
support or determine explosion in the context of the neutrino heating paradigm.  
These include 1) turbulence behind the stalled shock wave; 2) 
the progenitor density structure, in particular the magnitude 
of the discontinuity at the silicon/oxygen interface; 3) neutrino heating
by inelastic neutrino-electron and neutrino-nucleon scattering; 4) many-body 
corrections to neutrino-nucleon scattering rates \citep{burrows:18,horowitz:17}; and 5) the magnitude and
distribution of seed perturbations \citep{CoOt13,couch:15,mueller:17} 
(though not explored here). We have found in this paper that most low-mass 
progenitor models explode in 3D, but that there may be a mass gap for 
the current generation of progenitor models near 12$-$14 M$_{\odot}$
where explosion is a bit more problematic.  This does not mean that these
stars don't explode, merely that there is a hint from our simulation experience
using state-of-the-art tools and modern stellar progenitors that non-rotating
cores in this intermediate mass range are less explodable. In addition, we
see in our 3D simulations, as with the previous generation of 2D simulations, that
simultaneous explosion and accretion (in different solid-angle sectors at a given time) 
helps maintain explosion in its early phases in some directions by continuing to 
power the emergent luminosity with an accretion component in other directions.  This 
can't happen in 1D, but with such a broken symmetry in multi-D it can. Related 
to this is a wasp-waist pinched structure seen in some debris
distributions and the strong dipolar character to many ejecta fields.
Moreover, and again \citep{burrows:18}, we see that compactness parameter 
in and of itself is not predictive of explosion.

Though the 3D calculations in this new tranche of full-physics 
supernova models suggest that we and the community of supernova 
theorists have reached an important milestone in the theory of
core-collapse supernovae with the routine generation of sophisticated 
3D explosion simulations, there are a number of caveats that bear 
listing. First, though not outdone by any extant 
full-physics 3D runs, the spatial and energy-group resolution
of our simulations should be improved $-$ a resolution study is in order.
Second, we have implemented and fielded with F{\sc{ornax}}
an approximate GR variant, wherein the correction to the monopolar 
strength of gravity has been addressed, along with the gravitational 
redshift of the neutrinos, but bonafide GR \citep{2016ApJS..222...20K,ott2018_rel}
has been left to later development. Third, we have employed a many-body
correction to the axial-vector coupling term in the neutrino-nucleon 
scattering rates that is still approximate  \citep{horowitz:17} and 
have not incorporated the still-uncalculated corresponding correction to the 
charged-current absorption rates \citep{sawyer1999,roberts2012,roberts_reddy2017}.
Since we have here and elsewhere \citep{burrows:18,vartanyan2018a} found such effects to be 
important, this remains an important topic for future investigation. Fourth, for this study
we have employed the SFHo EOS and though competitive it is not definitive.
The dependence of the outcome of stellar collapse on the EOS remains to be determined 
in full.  Fifth, though we perform truly 3D neutrino transfer solving for two angular 
moments of the specific intensity with a vector neutrino flux, this is not multi-angle
transport.  The assumed tensor closure form is an ansatz and the higher-order
moments (second and third) are provided analytically \citep{vaytet:11}. 
Though for pseudo-spherical problems such an approach can be quite accurate 
\citep{richers_nagakura2017}, ultimately full Boltzmann transport in seven dimensions 
(three space, three momentum space, and time), currently too expensive, will 
be necessary. In addition, it may well be that aspects of neutrino oscillations 
must be factored in, raising yet further the ultimate necessary level of 
computational complexity. Sixth, the mapping between progenitor 
structure and ZAMS mass is still in flux. Though palpable and 
enduring progress has been made these last few decades in determining 
the character of massive star evolution, since the density, seed perturbation, 
and rotational profiles of the collapsing cores of massive stars 
are not yet reliably determined in detail, one should be cautious in 
declaring what a given mass star might do at the end of its life.
In addition, we note that core rotation can affect the outcomes, but that
rapid rotation that results in neutron-star spin periods less $\sim$50 milliseconds
should be rare \citep{faucher_kaspi}. Such a rotation period is considered
``slow" by CCSN theorists. Nevertheless, there is such a densely-packed range in 
published massive-star structures for the full range of ZAMS masses (see Figure \ref{profiles}) that we find it difficult 
to conclude that what is now in the literature does not span a realistic range of progenitor
structures.  However, and importantly, chaos and stochasticity in the turbulent 
pre-supernova phases will translate into a spread of outcomes in all the astronomical observables 
(e.g., explosion energies, kick velocities, final neutron star spins and masses,
nucleosynthesis, debris spatial distributions). What these spreads are   
is currently unknown. 

Hence, there still remains much to do in CCSN theory
that will keep investigators busy for many years.  Nevertheless, with the recent
emergence of sophisticated codes such as F{\sc{ornax}} to generate multiple 3D simulations every year,
we have entered a new era in the study of supernova explosions.  No longer is the community limited
to one or a few expensive 3D runs per year whose individual import is ambiguous.  Now, with
adequate computational resources many, many full-physics 3D simulations per year are possible.  
This enables the broader exploration of parameter space and can lead to a more forgiving 
theoretical environment. The few inevitable mistakes necessary to make real progress in 
a science are no longer as dire. The last two decades marked the era of commodity 2D simulations
that facilitated real progress in parameter exploration and constituted a palpable leap
in overall understanding.  After $\sim$50 years of development in physics, technique,
and computational capabilities, we have now entered the corresponding era in 3D modeling.
As a result, with good reason can we expect in the years to come, further (and final?) 
leaps in insight into this, one of the last remaining fundamental theoretical 
challenges of stellar astrophysics.

\section*{Acknowledgements}

The authors acknowledge fundamental contributions to 
this effort by Josh Dolence and Aaron Skinner.
We also acknowledge help with visualization using VisIt from 
Viktoriya Morozova, fruitful conversations with Hiroki 
Nagakura and Sean Couch, Evan O'Connor regarding the equation of state, 
Gabriel Mart\'inez-Pinedo concerning electron capture on heavy nuclei, 
Tug Sukhbold and Stan Woosley for providing details concerning the 
initial models, and Todd Thompson regarding inelastic scattering. We acknowledge support 
from the U.S. Department of Energy Office of Science and the Office 
of Advanced Scientific Computing Research via the Scientific Discovery 
through Advanced Computing (SciDAC4) program and Grant DE-SC0018297 
(subaward 00009650). In addition, we gratefully acknowledge support 
from the U.S. NSF under Grants AST-1714267 and PHY-1144374 (the latter 
via the Max-Planck/Princeton Center (MPPC) for Plasma Physics). DR 
acknowledges partial support as a Frank and Peggy Taplin Fellow at 
the Institute for Advanced Study. This overall research project is 
part of the Blue Waters sustained-petascale computing project, 
which is supported by the National Science Foundation (awards OCI-0725070 
and ACI-1238993) and the state of Illinois. Blue Waters is a joint effort 
of the University of Illinois at Urbana-Champaign and its National Center 
for Supercomputing Applications. This general project is also part of 
the ``Three-Dimensional Simulations of Core-Collapse Supernovae" PRAC 
allocation support by the National Science Foundation (under award \#OAC-1809073).  
Moreover, access under the local award \#TG-AST170045 
to the resource Stampede2 in the Extreme Science and Engineering Discovery 
Environment (XSEDE), which is supported by National Science Foundation grant 
number ACI-1548562, was crucial to the completion of this work.  Finally, 
the authors employed computational resources provided by the TIGRESS high 
performance computer center at Princeton University, which is jointly 
supported by the Princeton Institute for Computational Science and 
Engineering (PICSciE) and the Princeton University Office of Information 
Technology, and acknowledge our continuing allocation at the National 
Energy Research Scientific Computing Center (NERSC), which is 
supported by the Office of Science of the US Department of Energy
(DOE) under contract DE-AC03-76SF00098.

\onecolumn

\begin{table}
  \center{
  \begin{tabular}{lllllll}
    \hline
           & t(final)    & Exp. Energy &  NS mass (bar.) & NS mass (grav.) & PNS Radius & Env. Binding \\

           & (s)         & ($10^{50}$ ergs) & (M$_{\odot}$) & (M$_{\odot}$) & (km) & ($10^{50}$ ergs) \\
    \hline
    s9.0-2D   & 1.41  & 0.71      & 1.358 & 1.246 & 21.3  & -0.0206   \\
    s9.0-3D   & 1.042 & 1.02      & 1.342 & 1.233 & 23.5  &  $-$      \\
    s10.0-2D  & 1.41  & 0.62      & 1.524 & 1.382 & 20.8  & -0.0953   \\
    s10.0-3D  & 0.767 & 0.21      & 1.495 & 1.358 & 26.3  &  $-$      \\
    s11.0-2D  & 1.41  & 0.78      & 1.482 & 1.348 & 20.8  & -0.17     \\
    s11.0-3D  & 0.568 & 0.75      & 1.444 & 1.317 & 29.7  &  $-$      \\
    s12.0-2D  & 1.41  & 1.55      & 1.568 & 1.417 & 20.8  & -0.33     \\
    s12.0-3D  & 0.694 & 0.55      & 1.507 & 1.369 & 28.0  &  $-$      \\
    s13.0-2D  & 1.311 & -0.48     & 1.854 & 1.642 & 21.9  & -0.48          \\
    s13.0-3D  & 0.674 & -0.48     & 1.752 & 1.564 & 29.1  &  $-$      \\
    \hline
  \end{tabular}
  \caption{Some basic model results for the collection of 3D and 2D models calculated
for this paper.  The model names are followed by the post-bounce time at the end of each simulation,
the total explosion energy at the end of each run, the associated ``final" baryonic and gravitational masses 
and proto-neutron-star (PNS) radii (average radius of the 10$^{11}$ gm cm $^{-3}$ surface), and the binding
energy of the off-grid stellar envelope. Note that the explosion energies have been corrected for the latter
and that after correction all the exploding 3D models have positive explosion energies. The exception
is the 13-M$_{\odot}$ model, which has yet to explode by the end of both our 2D and 3D simulations.}
  \label{sn_tab}
  }
\end{table}

\begin{table}
  \center{
  \begin{tabular}{llll}
    \hline
           & t(final)    & Mean Shock Radius &  Mean Shock Speed \\

           & (s)         & (1000 km) & (1000 km s$^{-1}$)  \\
    \hline
    s9.0-2D   & 1.41  & 15.24    & 14.19    \\
    s9.0-3D   & 1.042 & 12.42    & 16.29    \\
    s10.0-2D  & 1.41  & 7.70     & 10.62    \\
    s10.0-3D  & 0.767 & 1.96     & 6.65     \\
    s11.0-2D  & 1.41  & 9.18     & 7.41     \\
    s11.0-3D  & 0.568 & 2.75     & 8.00     \\
    s12.0-2D  & 1.41  & 8.72     & 8.08     \\
    s12.0-3D  & 0.694 & 2.66    & 6.85     \\
    s13.0-2D  & 1.311 & 0.06     & 0.067    \\
    s13.0-3D  & 0.674 & 0.09     & 0.048    \\
    \hline
  \end{tabular}
  \caption{For the runs presented in this paper, the mean shock radius (in units of 1000 kilometers) 
and mean shock speed (in units of 1000 km s$^{-1}$) at the end of each simulation. Note that 
the shock is still stalled at the end of the simulation only 
for the 2D and 3D 13-M$_{\odot}$ models.}
  \label{sn_tab2}
  }
\end{table}

\bibliographystyle{mnras}
\bibliography{sn}

\end{document}